\documentclass[twocolumn,prl,superscriptaddress,noshowpacs]{revtex4-2}  

\usepackage{graphicx,nicefrac}
\usepackage{amssymb}
\usepackage{amsfonts}
\usepackage{xcolor}
\usepackage{amsmath}
\usepackage{siunitx}
\usepackage{soul}
\usepackage{flushend}

\usepackage[utf8]{inputenc}

\usepackage{hyperref}

\usepackage{amsmath}
\usepackage{mathtools}
\usepackage{braket}
\usepackage{dsfont}
\usepackage{bigints}
\usepackage{graphicx}
\usepackage{amssymb}
\usepackage{xcolor}
\hypersetup{
	colorlinks,
	linkcolor={red!95!black},
	citecolor={green!60!black},
	urlcolor={blue!95!black}
}

\usepackage{bbold}

\usepackage{soul}

\graphicspath{ {./Images/} }

\newcommand{\prlsection}[1]{\paragraph{\textbf {\textit {#1}} ---}}
\newcommand{\cll}{\text{$\chi$LL~}}

\newcommand{\ggtrapprox}{\mathrel{\vcenter{
			\offinterlineskip\halign{\hfil$##$\hfil\cr
				\gg\cr\noalign{\kern1pt}\sim\cr\noalign{\kern-1pt}}}}}

\newcommand{\approxpropto}{\mathrel{\vcenter{
			\offinterlineskip\halign{\hfil$##$\cr
				\propto\cr\noalign{\kern2pt}\sim\cr\noalign{\kern-2pt}}}}}

\newcommand{\nodagger}{{\vphantom{\dagger}}}

\usepackage{comment}
\usepackage{environ}

\newif\ifSM

\SMtrue

\ifSM
\NewEnviron{supplementalMaterials}{%
	\clearpage
	\setcounter{equation}{0}
	\setcounter{page}{0}	
	\BODY
}
\else
\excludecomment{supplementalMaterials}
\fi

\makeatletter
\def\maketitle{
	\@author@finish
	\title@column\titleblock@produce
	\suppressfloats[t]}
\makeatother

\usepackage{bbold}
\usepackage{soul}

\graphicspath{ {./Images/} }

\begin{document}

\title{
	Quantum nonlinear optics on the edge of {a} few{-}particle fractional quantum Hall {fluid} in {a small lattice}
}

\author{Alberto Nardin}
\affiliation{Universit\'e Paris-Saclay, CNRS, LPTMS, 91405 Orsay, France}
\email{alberto.nardin@universite-paris-saclay.fr}

\author{Daniele De Bernardis}
\affiliation{INO-CNR Pitaevskii BEC Center and Dipartimento di Fisica, Universit{\`a} di Trento, 38123 Povo, Italy}
\affiliation{National Institute of Optics [Consiglio Nazionale delle Ricerche (CNR)–INO], care of European Laboratory for Non-Linear
	Spectroscopy (LENS), Via Nello Carrara 1, Sesto Fiorentino 50019, Italy}
\email{daniele.debernardis@ino.it}

\author{Rifat Onur Umucal\i lar}
\affiliation{Department of Physics, Mimar Sinan Fine Arts University, 34380 Şişli, Istanbul, Türkiye}

\author{Leonardo Mazza}
\affiliation{Universit\'e Paris-Saclay, CNRS, LPTMS, 91405 Orsay, France}

\author{Matteo Rizzi}
\affiliation{Institute for Theoretical Physics, University of Cologne, D-50937 K\"oln, Germany}
\affiliation{Forschungszentrum Jülich GmbH, Institute of Quantum Control,\\
	Peter Gr\"unberg Institut (PGI-8), 52425 J\"ulich, Germany}

\author{Iacopo Carusotto}
\affiliation{INO-CNR Pitaevskii BEC Center and Dipartimento di Fisica, Universit{\`a} di Trento, 38123 Povo, Italy}

\begin{abstract}
	We study the  
	quantum dynamics {in response to time-dependent external potentials} of the edge {modes} of {a} small fractional quantum Hall {fluid} {composed of few particles on a lattice} {in a bosonic Laughlin-like state at filling $\nu=1/2$}.
	We show that the nonlinear chiral Luttinger liquid theory provides a quantitatively accurate description even for the small {lattices} that are available in state-of-the-art experiments{, away} from the continuum limit. 
	Experimentally{-}accessible {data related to} the quantized value of the bulk transverse Hall conductivity are identified both in the linear and the non-linear response to an external excitation.
	The strong nonlinearity induced by the open boundaries is responsible for sizable quantum blockade effects, leading to the generation of nonclassical states of the edge modes.
\end{abstract}

\date{\today}

\maketitle

\prlsection{Introduction}
Fractional quantum Hall (FQH) liquids {\color{black}have attracted} continuous experimental and theoretical attention due to the remarkable interplay between topology and {\color{black}interparticle} correlations, featuring a robust quantization of the transverse Hall conductivity~\cite{von_Klitzing_Nature_Review2020},
{\color{black}collective fractional excitations 
	in the bulk}~\cite{Laughlin_PRL_1983,Arovas_PRL_1984,Nayak_RMP_2008, Nardin2023SpinStatistics}
and gapless chiral modes on its edge~\cite{Wen_PRL_1990,Wen_PRB_1990b,Wen_PRB_1991, Wen_PRB_1991b, Wen_AdvPhys_1995,Wen_intJModPhysB_1992,Chang_RMP_2003}. 
At low energies, these latter are well described in terms of an effective one-dimensional chiral Luttinger liquid ($\chi$LL) theory~\cite{Wen_AdvPhys_1995} and provide a powerful probe of the topological properties of the fluid~\cite{Wen_science_2019, dePicciotto_nat_1997,Chang_PRL_1996, Chang_RMP_2003,Banerjee_Nat_2018,Umansky_NatPhys_2023,Nakamura_Nature_2020,Bartolomei_science_2020, Nakamura_PRX_2023}.

Beyond electronic systems, a strong experimental attention is presently being devoted to the realization of FQH liquids in synthetic matter systems such as ultra-cold atoms under synthetic magnetic fields~\cite{Cooper_2008,Bloch_RMP_2008,Cooper_RMP_2019,Goldman_RepProgPhys_2014} or strongly interacting photons in nonlinear topological photonic devices~\cite{Carusotto_NatPhys_2020,Carusotto_RMP_2013,Ozawa_RMP_2019}. 
First experimental observations of small FQH clouds have been recently reported in both the atomic~\cite{Gemelke_2010,Leonard_Science_2023,Lunt_arXiv_2024} and photonic~\cite{Clark_Nat_2020,Wang_Science_2024} contexts.
Concurrently, a number of theoretical studies have proposed protocols that exploit the new manipulation and diagnostic tools that are available for these systems to obtain insight on both the bulk~\cite{Paredes_PRL_2001,Sorensen_PRL_2005,Hafezi_PRA_2007, Kapit_PRL_2012,Price2012Mapping, Cooper2013Reaching, Cooper_PRL_2015, Tran_SciAdv_2017,Raciunas_PRA_2018, Wang2018Floquet,  Umucalilar2018Timeofflight, Asteria_NatPhys_2019, Repellin_PRL_2019, Macaluso2019FusionChannels, Macaluso_PRR_2020, Repellin_PRA_2020, wang_2022_measurable, palm_2022_snapshot, Comparin2022Measurable,  Umucalilar_PRA_2023, Cr_pel_2024, palm_2024_growing} and the edge~\cite{Kjall_PRB_2012,Luo_PRB_2013, Dong2018Charge, Binanti_PRR_2024, redon2023realizing, nardin2023bisognanowichmann} physics.
However, state-of-the-art lattice setups~\cite{Leonard_Science_2023,Wang_Science_2024} are small,
the spatial confinement provided by open boundaries is not smooth, and the magnetic field too large for a continuum description to be accurate. One may therefore be concerned that the basic FQH physics is distorted and the topological features washed out.

In this Letter we show that the \cll nature of the FQH edge excitations is robust and unambiguously visible even in {a small lattice with few-particles, by studying the case of a bosonic Laughlin-like state at filling $\nu=1/2$}.
Turning the small spatial size and the open boundaries of state-of-the-art systems into an advantage, we anticipate that the non-linearity of the \cll induced by the spatial confinement~\cite{Fern_PRB_2017,Macaluso_PRA_2017,Macaluso_PRA_2018,Nardin_PRA_2023, nardin2023refermionized} mediates efficient three-wave mixing process between Luttinger quanta in the discrete chiral edge modes around the fluid. This extends to the FQH edge {the} quantum blockade effects which are well-known in electronic~\cite{Kastner:RMP1992}, photonic~\cite{Imamoglu_PRL_1997,Verger:PRB2006,birnbaum2005photon,Lang:PRL2011}, and opto-mechanical~\cite{Rabl:PRL11,Girvin:PRL11} systems and opens a path towards the generation and control of nonclassical states of the edge modes.

\prlsection{Model}
\begin{figure}
	\centering
	\includegraphics[width=\columnwidth]{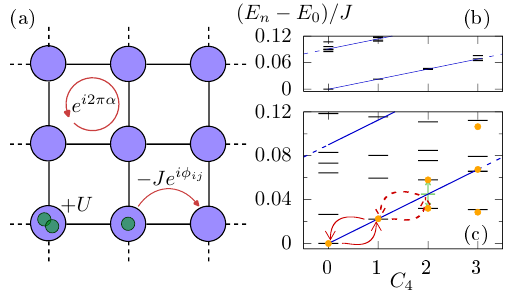}
	\vspace{-0.5cm}
	\caption{(a) Scheme of the model under consideration. 
		(b-c) Numerically calculated spectrum of the many-body eigenstates for $N=4$ atoms with $\alpha=0.15$ in (b) $12\times12$ and (c) $10\times10$ square lattices (black horizontal lines). 
		In panel (c), the confinement comes from the open boundaries, while in (b) it is dominated by an additional harmonic potential $V_h(\mathbf{r})=\omega_h^2 \,(\mathbf{r}-\mathbf{r}_0)^2/2$ of strength $\omega_h=0.14J$ around the lattice center $\mathbf{r}_0$.
		Comparison with the linear dispersion of a $\chi$LL  theory (blue dashed line) highlights the stronger  anharmonicity of the open boundary case of (c): here, the nonlinearity-induced splitting of the $l=2$ manifold is indicated by the green arrows, while the blockade effect is illustrated by the red full/dashed arrows indicating the resonant/detuned transitions.
		Orange points show the eigenenergies obtained from the diagonalization of the nonlinear \cll Hamiltonian in Eq.~\eqref{eq:NLchiLL}, {\color{black}where the parameters $\Omega,\Gamma$ are obtained from the energy of the lowest $C_4=1$ state and from the splitting of the two lowest $C_4=2$ states.}
	}
	\label{fig:model}
\end{figure}

We consider a system of $N$ bosonic particles moving in a $M\times M$ two-dimensional square lattice with open boundaries and lattice spacing $a$, pierced by a uniform and orthogonal (synthetic) magnetic field $B$, see Fig.~\ref{fig:model}(a).
We describe the system with the Hofstadter-Bose-Hubbard Hamiltonian~\cite{Sorensen_PRL_2005,Hafezi_PRA_2007,Gerster_PRB_2017} 
\begin{equation}
	\hat{H}_\text{HBH} = - J\sum_{\braket{ij}} \left( e^{i \phi_{ij}}\hat{a}_j^{\dag} \hat{a}^{\nodagger}_i + {\rm h.c.} \right) + \frac{U}{2}\sum_i \hat{a}_i^{\dag} \hat{a}_i^{\dag} \hat{a}^{\nodagger}_i \hat{a}^{\nodagger}_i
	\label{eq:modelH}
\end{equation}
where $\hat{a}^{\vphantom{\dagger}}_i$ ($\hat{a}^\dagger_i)$ are the destruction (creation) operators for a particle at site $i$.
{\color{black}The on-site interaction energy $U$ and the hopping strength $J>0$ are taken in the $U\gg J$ hard-core limit, but we numerically checked that our results do not change appreciably as long as $U\gtrsim4J$.}
The hopping phase $\phi_{ij}$ from site $i$ to $j$ is related to the magnetic vector potential through $\phi_{ij}=\frac{q}{\hbar}\int_{\mathbf{r}_i}^{\mathbf{r}_j} \mathbf{A}\cdot d\mathbf{r}$~\cite{hofstadter_PhysRevB.14.2239} and gives a gauge-invariant phase $\alpha=2\pi \phi/\phi_0$ {\color{black}when} hopping around a plaquette. Here, $\phi=B a^2$ is the magnetic flux piercing a plaquette and $\phi_0=h/q$ is the flux quantum. 
{\color{black}Our calculations are performed by exact-diagonalization methods, see the Supplemental Material (SM) for details~\cite{Note1}.}

This model is known to host a variety of FQH states~\cite{Hafezi_PRA_2007,Palm_PhysRevB.103.L161101,Moller_PRL_2009,Moller_PRL_2015,Boesl_PRB_2022}. In this Letter, we will focus on the simplest Laughlin $\nu=1/2$-like~\cite{Laughlin_PRL_1983} state that correspond to the ground state in the region $0.1\lesssim\alpha\lesssim0.2$ (indicated by the
transparency window in Fig.~\ref{fig:matrixElements}(a-d){\color{black}, corresponding to the region where the absolute value of the overlap between the ground state and the discretised Laughlin wavefunction exceeds $0.9$}).
The finite extension of the FQH region is allowed by the flexibility in the central density value offered by the density-depleted region at the boundary~\cite{Gerster_PRB_2017,Macaluso_PRR_2020,Repellin_PRA_2020}.

\begin{figure}[t]
	\centering
	\includegraphics[width=\columnwidth]{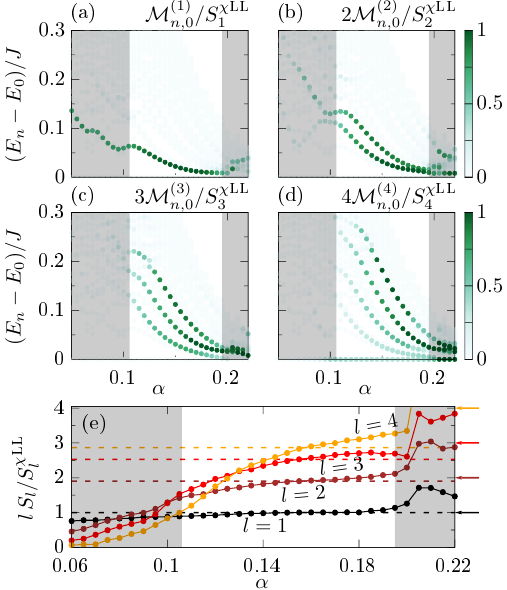}
	\vspace{-0.5cm}
	\caption{Magnetic field dependence of the excited state eigenenergies in the different $l=1,\ldots,4$ (a-d) sectors. Points  are coloured according to the normalized matrix element $ l \mathcal{M}_{n,0}^{(l)}/S_l$. 
		{\color{black}The white transparency window signals that the absolute value of the overlap between the ground state and the discretized Laughlin state is larger than $0.9$.}
		(e) Normalized static structure factor (full lines) compared to the \cll prediction in the thermodynamic limit (lateral arrows) and a finite-size calculation for a $N=4$ system in the continuum (horizontal dashed lines).    
		Same 
		parameters as in Fig.~\ref{fig:model}(c).}
	\label{fig:matrixElements}
\end{figure}

\prlsection{Nonlinear chiral Luttinger liquid}

In the basis labeled by the angular momentum $l>0$, the effective one-dimensional nonlinear $\cll$ theory~\cite{ImambekovGlazman_RMP_2012,Fern_PRB_2018,Nardin_PRA_2023} of the edge  of an anharmonically trapped FQH fluid reads
\begin{equation}
	\label{eq:NLchiLL}
	\hat{H} =\! \sum_{l>0} \Omega\, l\, \hat{b}_l^\dagger\hat{b}_l^\nodagger \,+\! \sum_{l,l'>0} \Gamma \sqrt{l l' (l+l')}\left[\hat{b}_{l+l'}^\dagger\hat{b}_{l}^\nodagger \hat{b}_{l'}^\nodagger \!+\! \textrm{h.c.}\right], 
\end{equation}
where $\hat{b}_{l}$ ($\hat{b}_{l}^\dagger$) are the bosonic {\color{black}annihilation} (creation) operators of Luttinger quanta in the $l$ mode. 
The first term describes the angular velocity $\Omega$ of chiral edge excitations, as determined by the spatial confinement. 
On top of this, the anharmonicity of the confinement gives the second term describing scattering processes between Luttinger quanta mediated by a nonlinear term of strength $\Gamma\geq0$~\cite{Nardin_PRA_2023}.
Higher-order {\color{black}dispersive and non-linear terms} can be safely neglected as they are irrelevant for the low-energy processes considered in this work~\cite{Fern_PRB_2018, Nardin_PRA_2023,nardin2023refermionized}. 

If the trapping is purely harmonic, the theory reduces to the standard \cll {\color{black}with $\Gamma=0$}~\cite{Cazalilla_PRA_2003}, where the excitations are non-interacting and have a perfectly linear dispersion.
As such, the eigenstates collapse on the $\omega_l=\Omega l$ values
and, at each $l$, have a high degeneracy given by the number $P_l$ of inequivalent integer partitions of $l$~\cite{Wen_AdvPhys_1995}, that is $P_1=1,\,P_2=2,\,P_3=3,\,P_4=5,\,\hdots$. 
Under a generic confinement, the massive degeneracy of the eigenstates at a given $l$ is lifted by the nonlinearity in \eqref{eq:NLchiLL}.

\prlsection{Spectrum of edge excitations}
In a square lattice the full rotational symmetry is not present; a proper choice of the gauge $\phi_{ij}$~\cite{Kjall_PRB_2012} ensures that the discrete rotation operator $e^{i \hat{L}_z \frac{\pi}{4}}$ commutes with the system's Hamiltonian~\eqref{eq:modelH}, so that the angular momentum is conserved modulo $4$.
Examples of numerically computed excited-state spectra are plotted in Fig.~\ref{fig:model}(b,c) as a function of the {\color{black}quasi-}angular momentum $C_4$.
{\color{black}In order to compare them with the spectrum of $\hat H$ in~\eqref{eq:NLchiLL}, we need to fold the latter with period 4.}
In Fig.~\ref{fig:model}(b) the trapping is dominated by {\color{black}an additional harmonic} confinement, so that the spectrum is almost linear and resembles the standard \cll prediction (blue dashed line) with minor {\color{black}corrections.}
In Fig.~\ref{fig:model}(c), instead, the only confinement is due to the open boundaries {\color{black}of the lattice}: its intrinsic anharmonicity is responsible for marked nonlinear effects and the wide splitting of the multiplets at given $l$. 
Quite remarkably, the numerical eigenenergies are accurately recovered by the predictions of the nonlinear \cll theory~\eqref{eq:NLchiLL}.

This physics is {\color{black}easiest understood} in the simplest $l\!\!=\!\!2$ manifold. 
From the form of the {\color{black}nonlinear \cll Hamiltonian} Eq.~\eqref{eq:NLchiLL}, it is immediate to recognize that the effect of the nonlinearity is to coherently convert a pair of $l=1$ Luttinger quanta into a single $l=2$ Luttinger quantum and viceversa, so that the eigenstates are non-classical superpositions of the $\hat{b}_{2}^\dagger|0\rangle$ and $(\hat{b}_{1}^\dagger)^2|0\rangle$ Fock states. 
In analogy with the biexciton Feshbach blockade effect~\cite{carusotto2010feshbach}, 
the anharmonicity of the resulting spectrum leads to a marked blockade effect indicated by the red arrows in the level scheme of Fig.~\ref{fig:model}(c): an external excitation that is resonant with the $l=1$ mode at linear regime will not be able to inject a second quantum, as {\color{black}all the relevant} eigenstates are shifted away in energy. 

\prlsection{Linear response}
We now proceed with a study of the response of the fluid to external perturbations, starting from the linear regime of a weak excitation~\cite{StringariPitaevskii2018BoseBook}.   
{\color{black}We consider a pulsed external potential of well-defined angular momentum $l$ 
	of the form
	$\hat{V}_l(t)=\sum_{i} V_l(\mathbf{r}_i,t) \hat{a}_i^\dagger \hat{a}_i$ with
	$V_l(\mathbf r,t) = \lambda_l \left( z^l e^{-i \omega_l t} + {\rm c.c.}\right) e^{- t^2 / \tau^2}$, where $\omega_l$ is the carrier frequency of the pulse, $\tau$ and $\lambda_l$ are its duration and strength, and  $z=x+i y$.}
Beyond their straightforward realization in synthetic matter systems~\cite{Goldman_PRL_2012,Nardin_PRA_2023,Binanti_PRR_2024}, such potentials can nowadays be implemented also for neutral edge excitations in electronic FQH systems~\cite{Yusa_PRR_2022}.

{\color{black}For each $l\geq1$, this potential generates a force with both radial and azimuthal components.
	This latter, due the transverse Hall response, induces a radial particle flow which is responsible for modifying the shape of the cloud.
	As we review in the SM~\cite{Note1}, simple arguments based on the hydrodynamics of 2D FQH fluids predict that the variation of the edge density in response to the applied potential is proportional to the quantized transverse Hall conductivity $\sigma_{xy}=\nu q^2/h$, in agreement with the $\cll$ theory~\cite{Wen_AdvPhys_1995}.
	On this basis, we expect that matrix elements of suitable perturbations will be related to $\sigma_{xy}$ also in our small lattice geometry, as we are now going to demonstrate.}

{\color{black}Within linear response~\cite{StringariPitaevskii2018BoseBook}, each} excited state $\ket{C_4,n}$ (i.e.~the $n$-th excited state at {\color{black}quasi-}angular momentum $C_4$), will give a peak in the frequency-dependent response, centred at $\hbar\omega_{C_4,n}=E_{C_4,n}-E_{0,0}$ ($E_{0,0}$ being the ground-state energy) and with a strength determined by the matrix element
$\mathcal{M}_{n,0}^{(l)} = \left|\Braket{C_4,n\left|\sum\nolimits_{i} z_i^l \hat{a}^\dagger_i \hat a_i^\nodagger\right|0,0}\right|^2$ with $C_4=l \pmod{4}$.
{\color{black}Note that only the positive-frequency part of the potential $V_l(\mathbf r, t)$ contributes to the response.} 
The frequencies and the strengths of the peaks for the lowest $l=1\ldots 4$ excitation channels are summarized in Figs.~\ref{fig:matrixElements}(a-d).  
In spite of the reduced rotational symmetry, the matrix element weight is concentrated only on a few eigenstates~\cite{Binanti_PRR_2024}.
{\color{black}
	The $l=1$ dipole mode that is visible in Fig.~\ref{fig:matrixElements}(a) for all values of $\alpha$ can be associated with the center-of-mass motion of the cloud, according to Kohn's theorem~\cite{Kohn_PR_1961} (see SM~\cite{Note1}). The $l\geq 2$ modes are instead sensitive to the many-body state and display the peculiar \cll structure and multiplicity in the FQH region. In particular, the fact that the number of visible $l=4$ states in Fig.~\ref{fig:matrixElements}(d) does not match the number of partitions $P_{l=4}=5$ confirms the presence of uncoupled  dark edge eigenstates, first predicted {\color{black}for a continuum model} by the fermionized theory of~\cite{nardin2023refermionized}.}

{\color{black}Summing over the lowest few hundred eigenstates of $\hat H_{\rm HBH}$ (all those we obtained through Lanczos-based methods~\cite{slepc})} at a given $C_4 = l \pmod{4}$ that are illustrated in Figs.~\ref{fig:matrixElements}(a-d) provides the numerical values for the structure factor of the edge  $S_l=\sum_n \mathcal{M}_{n,0}^{(l)}$ that we display in Fig.~\ref{fig:matrixElements}(e) as full circles.
These exact diagonalization results for the lattice geometry are compared to the numerical ones for a $N=4$ Laughlin state in a continuum geometry~\cite{Nardin_PRA_2023} (horizontal dashed lines) and to the \cll prediction in the thermodynamic limit $S^{\chi LL}_l=\nu l R_{\text{cl}}^{2l}$ (lateral arrows).
{\color{black}In this last expression, $R_{\text{cl}}=\sqrt{2N/\nu}\,l_B$ is the radius of the FQH cloud and $l_B=a/\sqrt{2\pi\alpha}$ is the magnetic length.}

For sufficiently small $l$, a clear plateau is visible in the region around $\alpha\sim0.15$ of good overlap with the discrete Laughlin state highlighted in Fig.~\ref{fig:matrixElements}(a-d). 
The plateaus {\color{black}of} the $l=1$ and $l=2$ modes are in perfect {and good} quantitative agreement with the \cll prediction{, respectively}. 
For $l>2$ the corrections due to the small number of particles induce sizeable deviations between the $N=4$ lattice numerics and \cll predictions, but remain in quantitative agreement with numerical results for a $N=4$ Laughlin state~\cite{Note1}.
This suggests that the observed deviations for the higher multipole excitations are mostly affected by the small particle number rather than by the lattice discretization.

\begin{figure*}
	\centering
	\includegraphics[width=\textwidth]{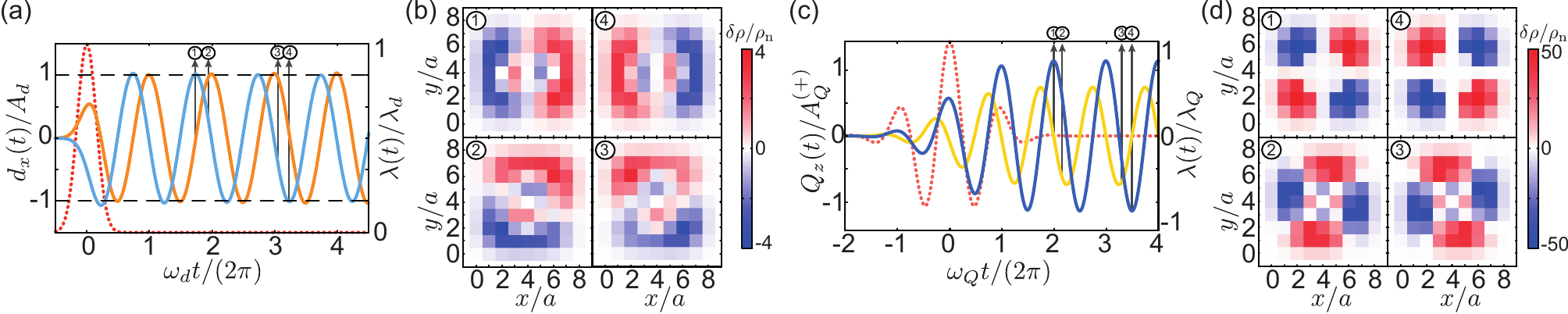}
	\vspace{-0.5cm}
	\caption{
		(a) Time evolution of the dipole moment $d_x(t)$ (blue solid line) and $d_y(t)$ (orange solid line). The pulsed excitation temporal profile $\lambda(t)$ is shown as a red dotted line.
		(b) Snapshots of the normalized density variation $\delta\rho/\rho_{\rm n}$ at the different time instants indicated in (a). We used $N=3$, $M=9$, $\alpha = 0.15$, $\lambda_1 = 10^{-4} J/a$, $\tau J/\hbar = 50$ and {\color{black}a normalization factor} $\rho_{\rm n} =  \lambda_1/(aJ)$. 
		The dipole frequency is $\hbar \omega_d /J\approx 0.027$.
		(c) Time evolution of the real ${\rm Re}[Q_z(t)]$ (yellow solid line) and imaginary ${\rm Im}[Q_z(t)]$ (blue solid line) parts of the quadrupole moment.
		The pulsed excitation temporal profile $\lambda(t)$ is shown as a red dotted line.
		(d) Snapshots of the normalized density variation $\delta\rho/\rho_{\rm n}$ at the different time instants indicated in (c). We used $N=3$, $M=9$, $\alpha = 0.15$, $\lambda_2 = 10^{-5} J/a^2$, $\tau J/\hbar = 160$ and  $\rho_{\rm n} =  \lambda_2/{\color{black}J}$. The quadrupole frequency is $\hbar \omega_2 /J\approx 0.035$.
	}
	\label{fig:dipole_quadrupole}
\end{figure*}

\prlsection{Measuring the transverse conductivity}
These numerical results show that the value of the edge structure factor $S_l$ matches the FQH {\cll} prediction also in a small lattice geometry {for low $l$}. As such they provide direct evidence of the quantized transverse Hall conductance and suggest that this latter can be experimentally assessed from a measurement of the response of the fluid to external potentials.

Fig.~\ref{fig:dipole_quadrupole}(a) illustrates this idea for the $l=1$ dipole {\color{black} mode, which}
can {\color{black} also (see SM~\cite{Note1})} be efficiently excited by means of a spatially uniform and temporally pulsed force, $V(\mathbf{r},t) = \lambda(t) x$ with $\lambda(t)={\color{black}\lambda_1} \exp \left[-(t/\tau)^2\right]$.
A few snapshots of the spatial profile of the density perturbation generated in the fluid at different times are shown in Fig.~\ref{fig:dipole_quadrupole}(b).  After the end of the perturbation, the $x,y$ components of the dipole moment [blue and orange lines in Fig.~\ref{fig:dipole_quadrupole}(a)] keep oscillating in quadrature at the natural frequency $\hbar\omega_d=E_{1,0}-E_{0,0}$ according to $d_{x}(t)=\Braket{\sum\nolimits_{i} x_i \hat{n}_i}(t)\simeq -A_d \sin\left(\omega_d t\right)$, {\color{black}$d_y(t)=A_d\cos(\omega_d t)$}. 
The amplitude {\color{black}can be computed using linear-response theory~\cite{Note1, StringariPitaevskii2018BoseBook} }, $A_d = \frac{S_1}{2\hbar}\,\widetilde{\lambda}(\omega_d)$,
proportional to  the Fourier component {\color{black}$\widetilde \lambda(\omega_d)$} of the excitation pulse at $\omega_d$. 
From a measurement of $d_{x}(t)$, it is then possible to estimate the structure factor $S_1$ and thus the transverse conductivity.

Application of a similar protocol to the higher $l>1$ modes requires isolating the several eigenstates forming the higher multiplets. 
The quadrupole $l=2$ case is illustrated in Fig.~\ref{fig:dipole_quadrupole}(c);
the two excited states {\color{black}at energies $E_{2,0}$ and $E_{2,1}$} can be separately addressed using a rotating saddle-shaped pulse of the form ${\color{black}V_2(\mathbf{r},t)}$, {\color{black}introduced in the previous section,} 
setting $\omega_2$ on resonance with the {\color{black} corresponding} transition.
{\color{black}In both cases,} at the end of the excitation pulse, the complex-valued quadrupole moment $Q_z(t) = \braket{\sum_i (x_i^2-y_i^2)\hat{a}^\dagger_i\hat a^\nodagger_i} + i \braket{\sum_i (2 x_i y_i)\hat{a}^\dagger_i \hat a^\nodagger_i}$ keeps oscillating 
as $Q_z(t)\simeq A_Q^{(+)} e^{i\omega_{Q} t}$ 
{\color{black}, where $\hbar \omega_Q = E_{2,n}-E_{0,0}$ with $n=0,1$ depending on the situation}. 
{\color{black}In the SM we discuss a small additional counter-rotating term that is responsible for the different amplitudes of the real and imaginary parts of $Q_z(t)$~\cite{Note1}.}
$A_Q^{(+)}$ is measurable 
and {\color{black}within linear response theory} gives direct information on the matrix element $\mathcal{M}_{n,0}^{(2)}=A_Q^{(+)} \frac{2 \hbar}{\sqrt{\pi}\tau \lambda_2}$. Upon summing over both quadrupole {\color{black}excitations}, the static structure factor $S_2$ can be extracted.

\begin{figure}
	\centering
	\includegraphics[width=\columnwidth]{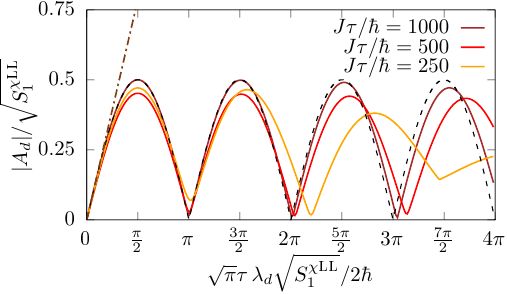}
	\vspace{-0.5cm}
	\caption{Amplitude of the dipole oscillations under the effect of a monochromatic excitation of different durations $\tau$ as indicated in the legend (solid red, brown, yellow lines) as a function of the (normalized) excitation strength $\lambda_d$. Dashed black line shows the
		theoretical form \eqref{eq:rabi} of the Rabi oscillations. 
		Same parameters as in Fig.~\ref{fig:dipole_quadrupole}.
		The brown dotted line is the linear regime prediction.} 
	\label{fig:Rabi}
\end{figure}

\prlsection{Quantum nonlinear dynamics}
The $\chi$LL theory becomes even richer when one goes beyond the linear response regime and considers excitations that are strong enough to induce nonlinear effects. These are mediated by the new interaction terms that appear in the Hamiltonian Eq.~\eqref{eq:NLchiLL} as a consequence of the anharmonic trapping potential. 
As a simple yet most illustrative example, we consider a quantum  blockade effect for edge modes. 
Let us consider a {\color{black}pulsed oscillating potential $V_1(\mathbf{r},t) = \lambda_1 \cos(\omega_d t) x\,e^{-t^2/\tau^2}$}, resonant with the dipole transition at frequency $\omega_d$ of duration $\tau$ and peak amplitude $\lambda_d$. Given the strong anharmonicity of the excitation spectrum highlighted in Fig.\ref{fig:model}(c), this perturbation is able to efficiently excite the $l=1$ excited state but remains well detuned from all two-excitation states in the higher $l=2$ manifold which therefore remain empty.
As a consequence, the dynamics recovers the one of a resonantly-driven two-level system
as first noticed in~\cite{Binanti_PRR_2024}, yet with the key advantage of the much stronger nonlinearity due to open boundaries.
At the end of the excitation sequence, each dipole component $d_{x,y}$ oscillates at the fast frequency $\omega_d$. The amplitude $A_d$ is determined by the periodic coherent transfer between the ground- and excited- state during the excitation sequence~\cite{CCT_atomPhotonInteractions},
at a Rabi frequency proportional to the perturbation strength $\lambda_d$.
As one can see in Fig.~\ref{fig:Rabi}, the Rabi oscillations 
are cleanest and closest to the theoretical prediction
\begin{equation}
	\label{eq:rabi}
	A_d = \frac{\sqrt{S_1}}{2}\sin\left(\sqrt{\pi}\tau\,\frac{\lambda_d\sqrt{S_1}}{2\hbar}\right)
\end{equation}
for longer-excitation times $\tau$, when the narrower spectrum of the excitation pulse makes the blockade effect on the $l=2$ manifold more effective. As typical in cavity quantum optics, a small system size is beneficial to the blockade effect, as it makes the dynamics faster by reinforcing the nonlinear splitting, and thus helps taming decoherence effects. 
Interestingly, the dependence of the Rabi oscillation frequency in Eq.~\eqref{eq:rabi} on the structure factor offers an alternative way to assess the quantized transverse conductivity of the FQH fluid.

Looking closer at the quantum state generated by the excitation sequence, we note that after a $2\pi$ (or multiple) pulse, the system is brought back to the ground state by the Rabi oscillation, while after a $\pi$ pulse the system is instead fully transferred to its $l=1$ excited state. In a quantum optical language~\cite{haroche_exploring_2013} this corresponds to the generation of a very non-classical Fock state with a single Luttinger quantum of excitation in the FQH edge, analogous to a single photon wavepacket in quantum optics. 
{\color{black}The generality of the blockade effect is confirmed in the SM~\cite{Note1}, where we show that it also occurs for an $l=2$ excitation.}

As another remarkable quantum effect, the spectral splitting of the $l=2$ response due to the nonlinear interaction terms of the \cll theory can be used to convert an $l=2$ Luttinger quantum into a pair of $l=1$ quanta so to obtain a source of correlated Luttinger pairs for quantum optical experiments on FQH edge modes.

\prlsection{Conclusions} 
In this work we have shown that the edge dynamics of {a} small lattice fractional quantum Hall {  bosonic Laughlin-like state at filling $\nu=1/2$}~\cite{Leonard_Science_2023,Wang_Science_2024} is well captured by a nonlinear chiral Luttinger liquid theory~\cite{Fern_PRB_2018,Nardin_PRA_2023}. 
{For large systems, the response to time-dependent external potentials 
	produces a signal that is expected to be quantitatively related to the
	quantized value of the transverse conductivity in the bulk; we have shown that also in the small lattices characterizing state-of-the-art setups~\cite{Clark_Nat_2020,Leonard_Science_2023,Wang_Science_2024,Lunt_arXiv_2024} it can be used to assess the topological nature of the boundary modes, provided one focuses on those at low angular momenta $l$.}
Turning their small spatial size and the open boundaries into an advantage, 
we have moved the first steps into quantum nonlinear optics of FQH edge modes: thanks to the intrinsic nonlinearity stemming from spatial confinement, quantum blockade effects lead to the generation of nonclassical states of the edge modes. 
Future theoretical work will deal with more complex quantum optical effects, such as the generation of entangled \cll pairs \cite{Anwar:Three-wave_review_2021}, solitonic and bunched excitations of the fractional quantum Hall edge~\cite{Bettelheim_PRL_2006,Wiegmann_PRL_2012,Mahmoodian:PhysRevX.10.031011} and exploitation of the edge 
as a information channel 
between localized impurities \cite{Simon:PhysRevLett.122.127701, debernardis:PRXQuantum.4.030306}.
{Finally, the extension of this work to other lattice fractional quantum Hall fluids~\cite{Moller_PRL_2009} is an exciting perspective, as different states might offer additional possibilities with respect to the case study proposed here.}

\prlsection{Acknowledgements}
We are thankful to E.~Macaluso for continuous collaboration in the first ground-breaking steps of this work, N.~Goldman, J.~Kwan, M.~Greiner and his group for insightful discussions.
A.N. acknowledges financial support from \textit{Fondazione Angelo dalla Riccia}, LoCoMacro 805252 from the European Research Council and LPTMS for warm hospitality. 
D.D.B. acknowledges funding from the European Union - NextGeneration EU, "Integrated infrastructure initiative in Photonic and Quantum Sciences" - I-PHOQS [IR0000016, ID D2B8D520, CUP B53C22001750006].
M.R. acknowledges support from the Deutsche Forschungsgemeinschaft (DFG) via project Grant No. 277101999 within the CRC network TR 183 (B01) and under Germany’s Excellence Strategy – Cluster
of Excellence Matter and Light for Quantum Computing
(ML4Q) EXC 2004/1 – 390534769.
I.C. acknowledges financial support from the European Union H2020-FETFLAG-2018-2020 project ``PhoQuS'' (n.820392), from the Provincia Autonoma di Trento, from the Q@TN initiative, and from the PNRR-MUR project PE0000023-NQSTI project, co-funded by the European Union - NextGeneration EU.
This work is part of HQI initiative (\texttt{www.hqi.fr}) and is supported by France 2030 under the French National Research Agency award number ``ANR-22-PNCQ-0002".

\let\oldaddcontentsline\addcontentsline
\renewcommand{\addcontentsline}[3]{}
	\bibliography{references}
	\bibliographystyle{mybibstyle}
\let\addcontentsline\oldaddcontentsline

\begin{supplementalMaterials}
	
	\title{Supplemental Material for\\ 	Quantum nonlinear optics on the edge of {a} few{-}particle fractional quantum Hall {fluid} in {a small lattice}}	
	
	\maketitle
	\onecolumngrid	
	{
		\hypersetup{linkcolor=black}
		\tableofcontents
	}
	
	\section{Kohn's theorem}
	In this section we briefly review the content of Kohn's theorem~\cite{Kohn_PR_1961}.
	The system one considers is a collection of $N$ quantum particles moving in two-dimensions under the influence of a perpendicular magnetic field, $B$, under the influence of internal forces alone and at most an external harmonic potential
	\begin{equation}
		H = \sum_{i=1}^N \frac{(\mathbf{p}_i-q\mathbf{A}(\mathbf{r}_i))^2}{2M} + \sum_{i=1}^N \frac{1}{2}M \omega^2 \mathbf{r}_i^2 + \sum_{1\leq i < j \leq N} V(\mathbf{r}_i-\mathbf{r}_j).
	\end{equation}
	We here use the symmetric gauge, $\mathbf{A}(\mathbf{r})=\frac{B}{2}(-y,x,0)$.
	The motion of the center of mass can be decoupled from the one of the $N-1$ remaining relative coordinates; 
	the former is ruled by a simple Hamiltonian: introducing the center of mass variable $\mathbf{R}=\frac{1}{N}\sum_i \mathbf{r}_i$ and its canonically conjugate momentum $\mathbf{P}=\sum_i \mathbf{p}_i$ one can separate the Hamiltonian $H=H_\text{rel}+H_\text{CM}$ into two parts, a part $H_\text{rel}$ which only depends on the $N-1$ relative coordinates and a second part $H_\text{CM}$ which instead depends only on the center of mass coordinates
	\begin{equation}
		\label{eq:CMHamiltonian}
		H_\text{CM}= \frac{\mathbf{P}^2}{2 M'} - \frac{\omega_c}{2} L_z 
		+ \frac{1}{2}M' \Omega^2\mathbf{R}^2,
	\end{equation}
	where $L_z = (\mathbf{R}\times \mathbf{P})\cdot \hat{z}$ is the center of mass angular momentum.
	Here we introduced the cyclotron frequency $\omega_c=\frac{qB}{M}$, 
	the total mass $M'=N M$
	and an effective oscillator frequency $\Omega=\sqrt{\omega^2 + \omega_c^2/4}$.
	The eigenfunctions are those of a harmonic oscillator; introducing radial $R$ and azimuthal $\theta$ coordinates for the center of mass position, these eigenfunctions can be written as
	\begin{equation}
		\label{eq:harmonicOscillatorWF}
		\Phi_{n,m}^{(\text{CM})}(R,\theta)=\frac{1}{\sqrt{\pi \frac{\Gamma\left(n+\frac{|m|+m}{2}+1\right)}{\Gamma\left(n-\frac{|m|-m}{2}+1\right)}\,l_\text{CM}^2}}\,\left(\frac{R}{l_\text{CM}}\right)^{|m|} e^{i m \theta}\,L_{n-\frac{|m|-m}{2}}^{(|m|)}\left(\frac{R^2}{l_\text{CM}^2}\right)\,e^{-\frac{R^2}{2l_\text{CM}^2}},
	\end{equation}
	where $\Gamma(n)$ is Euler's Gamma function, $L_n^{(\alpha)}(x)$ are associated Laguerre polynomials and we introduced a center of mass unit of length 
	$l_{CM}=\sqrt{\frac{\hbar}{M' \Omega}}$.
	These wavefunctions diagonalize the center of mass angular momentum, $L_z \Phi_{n,m}^{(\text{CM})}=m\,\Phi_{n,m}^{(\text{CM})}$, as well as the center of mass Hamiltonian Eq.~\eqref{eq:CMHamiltonian}, with eigenvalue
	\begin{equation}
		E_{n, m}^{(\text{CM})}=\frac{\hbar \Omega}{2}\,\left[4\left(n+\frac{m+1}{2}\right)-\frac{\omega_c}{\Omega}\,m\right].
	\end{equation}
	In terms of the principal quantum number $n\in\mathbb{N}$, the angular momentum quantum number $m$ must satisfy $m\geq-n$.
	For any value of $\omega$, the ground state can be seen to correspond to $n=0$, $m=0$.
	
	When the cyclotron frequency is much larger than the harmonic confinement one, $\omega_c\gg\omega$, the energies reduce to Landau level ones $E_{n, m}^{(\text{CM})}\simeq \hbar \omega_c\,\left(n+\frac{1}{2}\right)$, independent of the angular momentum quantum number $m$.
	This center of mass separation is relevant for the evaluation of the matrix elements of certain observables, such a
	\begin{equation}
		\begin{aligned}
			O =& \sum_i z_i = N R\, e^{+i\theta}
			\\
			O^* =& \sum_i z_i ^* = N R\, e^{-i\theta}
		\end{aligned}
	\end{equation}
	since they only depend on the center of mass variables and not on the $N-1$ relative coordinates. 
	In particular, $O$ couples the ground state only to $n=0$, $m=1$. This can be seen to be a low-energy excitation, since it does not change the Landau level quantum number $n$.
	$O^*$ on the other hand couples the ground state only to $n=1$, $m=-1$; it therefore represents an high-energy excitation (whose cost is set by the cyclotron energy $\hbar\omega_c$). This result is known as the Kohn's cyclotron's resonance~\cite{Kohn_PR_1961}.
	It is not difficult to show that
	\begin{equation}
		\bra{\Phi_{0,1}^{(\text{CM})}}O\ket{\Phi_{0,0}^{(\text{CM})}}=N\,l_\text{CM} \xrightarrow[\omega_c \gg \omega]{} \sqrt{2N}\,l_B 
	\end{equation}
	where $l_B=\sqrt{\frac{\hbar}{M \omega_c}}$. 
	Since this single state is contributing at low-energies (in the limit $\omega_c \gg \omega$), we get
	\begin{equation}
		S_1=|\bra{\Phi_{0,1}^{(\text{CM})}}O\ket{\Phi_{0,0}^{(\text{CM})}}|^2 = 2N\,l_B^2,
	\end{equation}
	which is the same result as $S_l = \nu l R_{\rm cl}^2$ presented in the main text as well as in the next section of this Supplemental Materials, in the case in which the angular momentum carried by the relevant excitation, $\sum_i z_i^l$, is $l=1$.
	Notice also that information on the relative motion will start to matter for any $l>1$.
	{\color{black}This shows how a single state coupled to the ground state through $O=\sum_i z_i$ can be expected to be present for any value of the synthetic magnetic field, both within and outside the fractional quantum Hall regime. 
		Even though one may be afraid that the lack of full translational symmetry invalidates the conclusions of these analytical considerations, our numerical results shown in  Fig~1(a) of the main text confirm that such a mode is indeed present except for the largest values of the synthetic magnetic field synthetic magnetic field shown.}
	
	\section{The Structure factor and the chiral Luttinger liquid theory}
	
	\subsection{Quantized transverse conductivity in the bulk and \cll commutation relations on the edge}
	In this section we want to give an alternative heuristic derivation of the commutation relations of the density operators of a \cll, which highlights the connection with the transverse conductivity of the bulk and does not rely on a specific form for the edge Hamiltonian.
	
	We consider a quantum Hall region infinite in the $\hat{x}$ direction and semi-infinite in the $\hat{y}$ direction, as we schematically depict in Fig.~\ref{fig:cll}. The magnetic field is along the $\hat{z}$ direction.
	If we apply a constant force in the $\hat{x}$ direction, by imposing an external potential $V=q E_x x$, a quantized current flows in the bulk $J_y=\sigma_{xy} E_x$ perpendicularly to the edge.
	
	Since charge is conserved, a continuity equation $\partial_t\rho=-\boldsymbol{\nabla}\cdot\boldsymbol{J}/q$ must hold. 
	For the present discussion, we only focus on the effect of this transverse current and not on the charge dynamics once the edge is reached (i.e. we neglect the chiral motion along the boundary caused by the sample confinement) by dropping the $\hat{x}$ contribution to the current; 
	more explicitly,
	$\partial_t\rho\simeq-\partial_yJ_y/q$.
	We now integrate this equation along the $\hat{y}$ direction, from the bulk region (where the density has not changed because of the bulk's incompressibility and $J_y^{\text{bulk}}=\sigma_{xy} E_x$) to the outside of the sample (where both the density and the current vanish)
	\begin{equation}
		\label{eq:dynamics1}
		\frac{\partial}{\partial t}\int dy \rho(x,y) = - \frac{1}{q}\int dy \frac{\partial J_y}{\partial y} = \frac{1}{q} J_y^{\text{bulk}}= \frac{\sigma_{xy}}{q^2} \frac{\partial V}{\partial x}.
	\end{equation}
	On the other hand, the left hand side of the expression can be rewritten by subtracting off the equilibrium density and $\int dy \left[\rho(x,y)-\rho_0(y)\right]=\delta\rho(x)$ can be seen as an effective edge density.
	
	\begin{figure}[t]
		\centering
		\includegraphics[width=12cm]{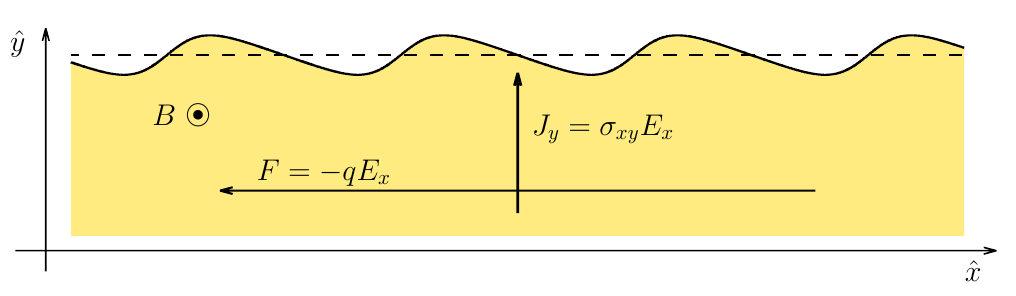}
		\caption{Schematic depiction of a quantum Hall system with an edge. The yellow region represents the incompressible quantum Hall fluid, whose edge (the black curly line) is deformed with respect to its ground-state position (the dashed black line).    
			A force is applied along the $\hat{x}$ direction and a current flows proportionally to the applied force in the $\hat{y}$ direction due to the presence of a strong perpendicular magnetic field.
			\label{fig:cll}}
	\end{figure}

	From a more microscopic perspective, $V$ couples to the system's density $\hat{\rho}=\hat{\Psi}^\dagger\hat{\Psi}^\nodagger$ through an interaction Hamiltonian
	\begin{equation}
		\label{eq:Hamiltonian}
		\hat{H}=\int dx \,dy \,V(x) (\hat{\rho} - \rho_0 ) = \int dx \,V(x) \int dy (\hat{\rho} - \rho_0 ) = \int dx \,V(x) \delta\hat{\rho},
	\end{equation}
	and Eq.~\eqref{eq:dynamics1} can be seen as the expectation value of a Heisenberg equation of motion
	\begin{equation}
		\label{eq:inflow}
		\frac{\partial}{\partial t}\delta\hat{\rho}(x) = \frac{\sigma_{xy}}{q^2} \frac{\partial V}{\partial x} + \hdots.
	\end{equation}
	{\color{black}Here, the ellipsis account for the free evolution of the edge-density $\delta\hat\rho(x)$.
		This equation crucially shows how a force parallel to the system's boundary $-\frac{\partial V}{\partial x}$ will induce a motion of particles from/towards the system's bulk, resulting in modulation of the edge-density $\delta\hat\rho(x)$ proportional to the transverse conductivity $\sigma_{xy}$.
	}
	
	In order for Eq.~\eqref{eq:inflow} to be compatible with the Hamiltonian Eq.~\eqref{eq:Hamiltonian}, we need
	\begin{equation}
		\frac{i}{\hbar}\left[\hat{H},\delta\hat{\rho}(x)\right] = \frac{\partial}{\partial t}\delta\hat{\rho}(x)=\frac{\sigma_{xy}}{q^2} \frac{\partial V}{\partial x},
	\end{equation}
	or 
	\begin{equation}
		\frac{i}{\hbar}\int dy\, V(y)\left[ \delta\hat{\rho}(y),\delta\hat{\rho}(x)\right]=\frac{\sigma_{xy}}{q^2} \frac{\partial V}{\partial x}.
	\end{equation}
	If we assume $\left[ \delta\hat{\rho}(y),\delta\hat{\rho}(x)\right] = \alpha \delta'(y-x)$, we get $\alpha = i \frac{\hbar \sigma_{xy}}{q^2}$ or
	\begin{equation}
		\label{eq:wen_commutator}
		\begin{split}
			\left[ \delta\hat{\rho}(y),\delta\hat{\rho}(x)\right] 
			=& i \frac{\sigma_{xy}}{\hbar q^2} \delta'(y-x) \\ 
			=& i \frac{\nu}{2\pi} \delta'(y-x)
		\end{split}
	\end{equation}
	which indeed does coincide with the famous \cll liquid commutation relations derived by Wen~\cite{Wen_PRL_1990,Wen_PRB_1990b,Wen_PRB_1991,Wen_intJModPhysB_1992,Wen_AdvPhys_1995}.
	This highlights the connection between the non-Fermi-liquid commutator Eq.~\eqref{eq:wen_commutator} and the bulk's transverse conductivity.
	{\color{black} Crucially, it is this commutator that it is responsible for the exact value of the structure factor that will be presented in the next section, Eq.~\eqref{eq:wen_SF}, highlighting the connection between the bulk's transverse conductivity and properties of the system's edge.}
	
	{\color{black}
		\subsection{ Structure factor calculation}}\label{Sec:Wen}
	\begin{figure}[t]
		\centering
		\includegraphics[width=12cm]{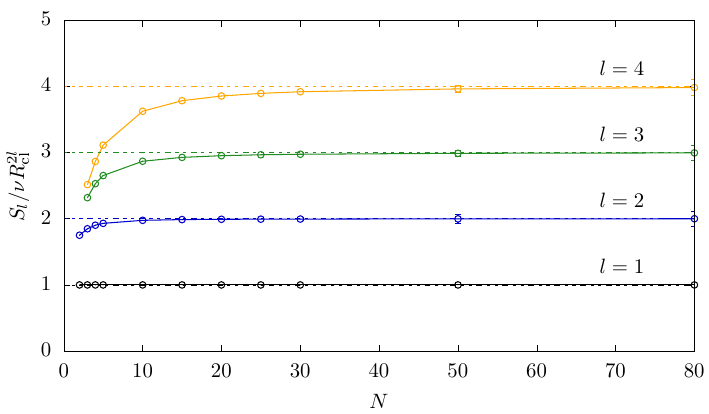}
		\caption{Low-energy static structure factor $S_l$ (see Eq.~\eqref{eq:ssf_rewriting}) for a continuum Laughlin state as a function of the number of particles $N$ {\color{black}up to $N=80$}. Our results are normalized to the expected result $S_l^\cll$ (see Eq.~\eqref{eq:wen_SF} per contributing state $l$, $S_l^\cll/l=\nu R_\text{cl}^{2l}$.
			The points are the results of Monte-Carlo simulations, while the dashed lines highlight the expected limit $l$.
			\label{fig:continuum}}
	\end{figure}
	
	In this section we review the theory for the static structure factor of the relevant observable $\sum_{i=1}^N z_i^l$ in the case of a $N$-particle Laughlin state in the continuum, in the thermodynamic limit $(N\gg 1)$. 
	In this case the system is fully rotationally symmetric; the angular momentum is a good quantum number, and we 
	can label the system's eigenstates as $\ket{l,n}$, with
	$l$ being the angular momentum with respect to the one of the Laughlin ground state. 
	Here $n$ is the additional quantum number which enumerates the system's eigenstates at fixed $l$.
	We are interested in the matrix elements
	\begin{equation}
		\label{eq:ssf}
		S_l = \sum_n \left|\Braket{l,n \left| \sum_{i=1}^N z_i^l \,\right| 0}\right|^2.
	\end{equation}
	These matrix elements can be conveniently rewritten as
	\begin{equation}
		\label{eq:ssf_rewriting}
		S_l = \sum_n \left|\int r^l e^{i l \theta} \Braket{l,n \left| \hat{\rho} (\mathbf{r}) \right|0} d^2\mathbf{r} \right|^2.
	\end{equation}
	Since the bulk is incompressible, the only contributions can come from a thin layer at the system's edge, close to system's classical radius $R_{\text{cl}}=\sqrt{2N/\nu} l_B$, $l_B$ being the magnetic length; in the large particle number ($N\gg 1$) limit the previous expression can be approximated by
	\begin{equation}
		\label{eq:approximated_SF}
		S_l \approx R_{\text{cl}}^{2l} \sum_n \left| \Braket{l,n \left| \widetilde{\rho}_l \right|0} \right|^2.
	\end{equation}
	where $\widetilde{\rho}_l =\int e^{i l \theta} \widetilde{\rho}(\theta)d\theta$ is the angular Fourier transform of a one-dimensional effective density $\widetilde{\rho}(\theta)= \int \hat{\rho}(\mathbf{r}) r dr$.
	Following Wen's bosonization
	we can replace the full $\widetilde{\rho}_l$ operators with bosonic fields
	\begin{equation}
		\label{eq:rho_l}
		\widetilde{\rho}_l = \sqrt{\nu l}\, \hat{b}^\dagger_l,
	\end{equation}
	where, {\color{black}as a consequence of Eq.~\eqref{eq:wen_commutator},} $\left[\hat{b}_l^{\vphantom{\dagger}}, \hat{b}_{l'}^{\dagger}\right]=\delta_{l,l'}$. We interpret the excited states $\ket{l,n}$ as being generated by the current algebra through
	\begin{equation}
		\label{eq:state}
		\ket{l,n}=\prod_{l'\in\boldsymbol{\lambda_l}}\frac{b^{\dagger}_{l'}}{\sqrt{n_{l'}!}}\ket{0}
	\end{equation}
	where $\boldsymbol{\lambda}_l$ is an integer partition with angular momentum $l$ and $n_{l'}$ the multiplicity with which $l'$ appears in $\boldsymbol{\lambda}_l$.
	The only state contributing to the summation in Eq.~\eqref{eq:approximated_SF} is therefore $\ket{l,n} = \hat{b}^\dagger_l\ket{0}$, and we get
	\begin{equation}
		\label{eq:wen_SF}
		S_l \approx \nu l\,R_{\text{cl}}^{2l}.
	\end{equation}
	{\color{black} Notice how the appearance of the (quantized) filling-fraction $\nu$ originates from the commutation relations Eq.~\eqref{eq:wen_commutator} obeyed by the edge-density operators. As we discussed in the previous section, this directly links $S_l$ to the transverse conductivity of the bulk.}
	{\color{black}In this last expression, the classical radius $R_{\text{cl}}$ of the FQH cloud accounts for the chosen $r^l$ radial dependence of the applied potential. 
	}
	
	Some numerical Monte-Carlo results obtained as detailed in~\cite{Nardin_PRA_2023,nardin2023refermionized} for the edge-excitations on top of a bosonic $\nu=1/2$ state are shown as a function of the number of bosons $N$ for $l=1,\hdots,4$ in Fig.~\ref{fig:continuum}.
	
	Notice that at $l=1$ the result is filling-fraction-independent and compatible with Kohn's theorem, as we detailed in the first section, and the numerical result for a translationally invariant system in Fig.~\ref{fig:continuum} indeed exactly matches the result of Eq.~\eqref{eq:wen_SF}; however it should also be noticed that there is no a-priori reason for our lattice system to obey Kohn's theorem and the result we show for $l=1$ has to be understood as a signature of fractional quantum Hall physics.

	\section{Dipolar excitation}
	
	\subsection{Perturbation theory}
	In this section we explicitly derive the system's dipole response to a dipolar pulsed excitation
	\begin{equation}
		\hat{V}(t) = \lambda(t) \sum_i x_i \hat{n}_i;
	\end{equation}
	we consider the pulse to have a Gaussian temporal profile $\lambda(t)=\lambda_d \exp(-t^2/\tau^2)$.
	We monitor the time evolution of the system's dipole moment
	\begin{equation}
		\hat{O}=\sum_i x_i \hat{n}_i.
	\end{equation}
	We consider our system to start in the ground state $\ket{0}$.
	
	To first order in perturbation theory we get
	\begin{equation}
		\label{eq:timeEvolutionObservable}
		\braket{\hat{O}}(t) \simeq  
		-\frac{2}{\hbar}\, \sum_n \text{Im} \left[
		\int_{-\infty}^t dt'\, \lambda(t') e^{-i \omega_{n,0} (t'-t)} |\braket{n|\hat{O}|0}|^2 \right]
	\end{equation}
	where $\omega_{n,0}=(E_n-E_0)/\hbar$ is the energy of the excited state $\ket{n}$. Here we used the fact that $\bra{0}\hat{O}\ket{0}=0$.
	
	In the Laughlin phase, at low energies a single mode alone contributes to the summation; namely the low-energy ``dipole" edge mode (see Fig.~1(a) in the main text).
	We define $\omega_d$ to be the frequency of such a transition. 
	Moreover, $|\braket{n|\hat{O}|0}|^2 = \left|\left<n\left|\sum_i\frac{x_i+i y_i}{2}\,\hat{n}_i\right|0\right>\right|^2=S_1/4$. The first equality is a consequence of the lattice fourfold rotational symmetry; the second one comes from the fact that the only contribution to the dipolar static structure factor is exhausted by this single matrix element (see Fig.~1(b,c) in the main text).
	
	If $t\gg\tau$ we can replace the upper integration limit in Eq.~\eqref{eq:timeEvolutionObservable} with $t\rightarrow\infty$, obtaining
	\begin{equation}
		\label{eq:timeEvolutionObservable_final}
		\braket{\hat{O}}(t) \simeq  
		-\frac{S_1}{2\hbar}\, \text{Im} \left[\widetilde{\lambda}(\omega_d) e^{i \omega_{d} t}\right]
	\end{equation}
	where we defined the Fourier transform $\widetilde{\lambda}(\omega_d)=\int_{-\infty}^\infty dt'\, \lambda(t') e^{-i \omega_{d} t'}=\lambda_d\tau\sqrt{\pi}\,\exp\left[-\left(\frac{\omega_d\tau}{2}\right)^2\right]$ of the pulsed excitation $\lambda(t)$.
	This is the equation we quoted in the main text.

	To conclude this section, let us notice that in order to control the strength of the perturbation $\tau$ cannot be chosen to be too large, otherwise the pulse becomes adiabatic with respect to the edge modes, as testified by the Fourier factor $\propto \tau \exp\left[-\left(\frac{\omega_d\tau}{2}\right)^2\right]$.
	This problem can be easily circumvented in principle by resonantly shaking the lattice with $\lambda(t)=\lambda_d \exp(-t^2/\tau^2)\cos(\Omega  t)$, $\Omega\approx\omega_d$. 
	Then, by neglecting the off-resonant term $\widetilde{\lambda}(\omega_d+\Omega)$, one obtains
	\begin{equation}
		\label{eq:timeEvolutionObservable_shaken}
		\braket{\hat{O}}(t) \simeq  
		-\frac{S_1}{4\hbar}\, \text{Im} \left[\widetilde{\lambda}(\omega_d-\Omega) e^{i \omega_{d} t}\right].
	\end{equation}
	Notice that all these results are valid only within the scope of linear perturbation theory.
	Non-linear effects are discussed in the next section in a two-level approximation.

	\subsection{Rabi oscillations}
	Thanks to the strong non-linearity imposed by the hard-wall condition, contrary to standard chiral Luttinger liquid (or in the presence of weak non-linearities) - the spectrum of the edge modes is not linearly dispersing with the excitation momentum.
	Highly non-linear behaviour can be seen in Fig.~1 in the main text.
	The idea is to exploit this non-linearity in order for the system to perform Rabi oscillations between the ground state and the dipole mode as if these two levels constituted an isolated two-level system.
	In this section we give some mathematical details on the performed calculations.
	
	In Heisenberg's representation $\ket{\widetilde{\psi}}=e^{i \hat{H}t/\hbar}\ket{\psi}$, Schr\"odinger equation reads
	\begin{equation}
		\label{eq:SchrodingerHeisenberg}
		i\hbar\frac{\partial\ket{\widetilde{\psi}}}{\partial t} = \widetilde{V}\ket{\widetilde{\psi}},
	\end{equation}
	where $\widetilde{V}=e^{i\hat{H}t/\hbar}\hat{V}e^{-i\hat{H}t/\hbar}$ is the excitation operator, $\hat{V}=\lambda(t)\hat{O}$. The observable we excite is again dipolar, $\hat{O}=\sum_i x_i \hat{n}_i$.
	As we briefly commented at the end of the previous section, we resonantly shake the lattice
	\begin{equation}
		\label{eq:resonant_shake}
		\lambda(t)=\lambda_d e^{-t^2/\tau^2}\,\cos(\Omega t).
	\end{equation}
	If the spectrum of edge modes is linearly dispersing with the momentum, as one would expect for smooth edges in the thermodynamic limit where the chiral Luttinger liquid description is expected to be valid, long excitations of this kind will not only drive a transition from the Laughlin ground state to its dipole moment, but also higher order excitations which on the long run drive the system out of these two levels.
	However, as we already anticipated, strong non-linearities imposed by the hard-wall confinement together with the small lattices analysed here make these higher-order transitions off-resonant. 
	We can therefore effectively describe the dynamics as being restricted to two levels, and we can expand $\ket{\widetilde{\psi}}=\widetilde{c}_0\ket{0}+\widetilde{c}_1\ket{1}$: the two states represent the ground state and the lowest-lying dipole state respectively, see Fig.~1 in the main text. 
	Taking the scalar products of Eq.~\eqref{eq:SchrodingerHeisenberg} with $\ket{0}$ and $\ket{1}$ we get
	\begin{equation}
		\begin{cases}
			i\hbar\frac{\partial\widetilde{c}_0}{\partial t} = \widetilde{c}_1 \bra{0}\widetilde{V}\ket{1}
			\\
			i\hbar\frac{\partial\widetilde{c}_1}{\partial t} = \widetilde{c}_0 \bra{1}\widetilde{V}\ket{0}.
		\end{cases}
	\end{equation}
	Here, we used the fact that $\bra{0}\hat{O}\ket{0}=\bra{1}\hat{O}\ket{1}=0$ by symmetry.
	On the other hand, $\bra{1}\hat{O}\ket{0}\equiv d_x=|d_x|e^{i\alpha}$ is non-zero. 
	The phase can be absorbed into the wavefunction amplitudes by defining $b_0=\widetilde{c}_0 e^{i\alpha/2}$, $b_1=\widetilde{c}_1 e^{-i\alpha/2}$.
	
	Since $\Omega\approx\omega_d$ we can neglect fast terms, obtaining
	\begin{equation}
		\begin{cases}
			i\frac{\partial{b}_0}{\partial t} = \Omega_R\, {b}_1(t)\,e^{i (\Omega-\omega_d)t} e^{-t^2/\tau^2}
			\\
			i\frac{\partial{b}_1}{\partial t} = \Omega_R\, {b}_0(t)\,e^{i (\omega_d-\Omega)t} e^{-t^2/\tau^2}.
		\end{cases}
	\end{equation}
	Here we defined an effective Rabi frequency
	\begin{equation}
		\Omega_R=\frac{\lambda_d|d_x|}{2\hbar}.
	\end{equation}
	For simplicity we consider the driving to be resonant, namely $\Omega=\omega_d$. The solutions with initial conditions $b_0(t\ll\tau)=1$ and $b_1(t\ll\tau)=0$ read
	\begin{align}
		\begin{cases}
			{b}_0(t)=\cos\left(\Lambda(t)\right)
			\\
			{b}_1(t)=-i\sin\left(\Lambda(t)\right)
		\end{cases}
		&\qquad
		\Lambda(t)=\frac{1}{2}\sqrt{\pi}\tau\,\Omega_R\left(1+\text{erf}\left(\frac{t}{\tau}\right)\right).
	\end{align}
	We can finally write down the time evolution of the dipole moment
	\begin{align}
		\braket{\hat{O}}(t)&=2\text{Re}\bigl(b_0(t)b_1^*(t)e^{i\omega_d t}|d_x|\bigr)=-|d_x|\sin(2\Lambda(t))\,\sin(\omega_d t).
	\end{align}
	As we explained in the previous section, $|d_x|=\sqrt{S_1}/2$.
	Therefore, at large times $t\gg\tau$ (since $\text{erf}(t/\tau)\rightarrow1$) we can write
	\begin{align}
		\braket{\hat{O}}(t\gg\tau)=-\frac{\sqrt{S_1}}{2}\sin\left(\sqrt{\pi}\tau\,\frac{\lambda_d\sqrt{S_1}}{2\hbar}\right)\,\sin(\omega_d t)
	\end{align}
	which is the result we quoted in the main text.
	Notice that the dipole oscillates as a pure sine-wave with an amplitude $\frac{\sqrt{S_1}}{2}\sin\left(\sqrt{\pi}\tau\,\frac{\lambda_d\sqrt{S_1}}{2\hbar}\right)$: this can be interpreted as the system performing Rabi oscillations with frequency $\Omega_R=\frac{\lambda_d \sqrt{S_1}}{\hbar}$, driven over an effective time-window $T=\sqrt{\pi}\tau$.
	
	Notice finally that when $\lambda_d\sqrt{S_1}\tau/\hbar\ll1$ it correctly recovers the perturbative result obtained in the previous section, Eq.~\eqref{eq:timeEvolutionObservable_shaken}.
	On the other hand, the strong anharmonicity due to the confinement allows one to get dipole responses {\color{black} which would otherwise not be possible by remaining in a perturbative regime}
	\begin{equation}
		|\braket{\hat{O}}|\approx \frac{\sqrt{S_1}}{2}
	\end{equation}
	provided one can perform a $\pi/2$ pulse which drives the system from its many-body ground state $\ket{0}$ to the coherent and equal superposition $(\ket{0}+\ket{1})/\sqrt{2}$; this in practice is limited by the pulse duration $\sqrt{\pi}\tau$, which is limited by the coherence time of the experiment.

	\begin{figure}[t]
		\centering
		\includegraphics[width=12cm]{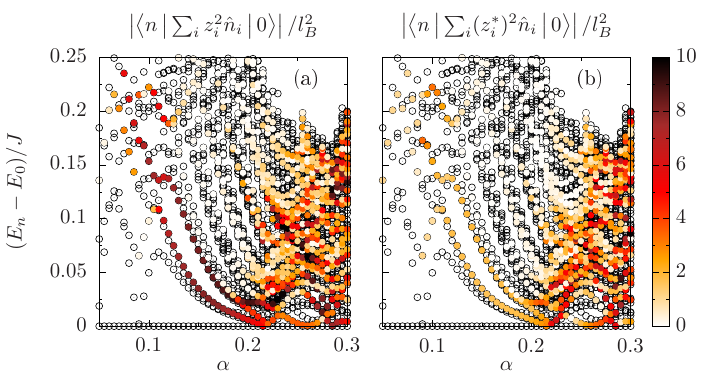}
		\caption{Energies (with respect to the ground  state one) for a system of $N=3$ bosons on a $9\times9$ lattice, as a function of the magnetic field per plaquette $\alpha$.
			The points have been coloured according to the value of (a) $\mathcal{M}_{n,0}^{(2)}=\left|\braket{n\left|\sum_i z_i^2 \hat{n}_i\right|0}\right|^2$ and (b) $\left|\braket{n\left|\sum_i (z_i^*)^2 \hat{n}_i\right|0}\right|^2$, {\color{black} i.e. the square modulus of the matrix elements of $\hat O_{z^2}$ and $\hat O_{{z^*}^2}$, respectively}.
			\label{fig:quadrupoles}}
	\end{figure}
	
	\begin{figure}[b]
		\centering
		\includegraphics[width=12cm]{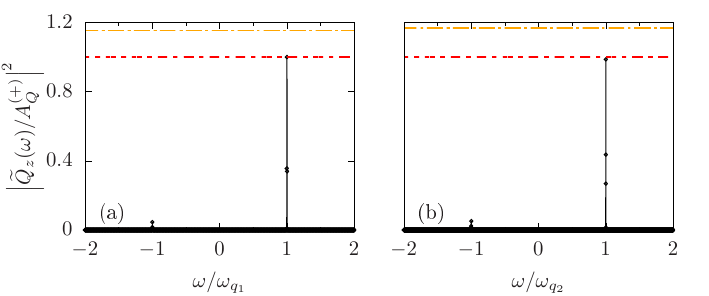}
		\caption{Square modulus of the Fourier transform $\tilde{Q}_z(\omega)$ of the quadrupole response $Q_z(t)$ at long times (after the pulsed excitation has been turned off), as a function of the frequency $\omega$ (normalized to the relevant quadrupolar frequencies (a) $\omega_{q_1}$ and (b) $\omega_{q_2}$.
			In both cases, two narrow peaks at $\pm1$ can be seen: Wen's sum rule Eq.~\eqref{eq:wen_SF} can be studied by analysing the positive frequency peaks in the two cases.
			The yellow horizontal line is the linear perturbation theory prediction (see Eq.~\eqref{eq:perturbativeQuadrupole}); the red one instead is the non-linear two-level system prediction (see Eq.~\eqref{eq:rabi_quadrupole}).
			{\color{black}The system consists here of $N=3$ hard-core bosons hopping on a $9\times9$ lattice, at synthetic magnetic flux per plaquette $\alpha=0.15$. The frequencies of the quadrupole modes are (a) $\omega_{q_1}\simeq=0.035J/\hbar$ and (b) $\omega_{q_2}=0.071J/\hbar$. A weak ($\lambda_Q= 2.2\times 10^{-5} J/a^2$) rotating saddle excitation (see Eq.~\eqref{eq:rotating_saddle}) has been pulsed over a long time-window $\tau=2000 \hbar/J$, which slightly drives the system out of the perturbative regime (yellow dashed-dotted line); the Rabi angles $\Theta_R=\Omega_R T$ (see Eq.~\eqref{eq:rabi_quadrupole}) are indeed (a) $\Theta_R\simeq0.65$ and (b) $\Theta_R\simeq0.68$.}
			\label{fig:quadrupole_powerSpectrum}}
	\end{figure}
	
	\section{Quadrupolar excitation}
	
	\subsection{Perturbation theory}
	Since the non-linearities introduced by the hard-wall confinement strongly split the two quadrupolar edge-modes (see Fig.~1(c) in the main text), the quadrupolar static structure factor is most easily measured by separately addressing the two transitions so as to measure the two transition matrix elements.
	In this section, we label these two states as $\ket{q_1}$ and $\ket{q_2}$.
	
	We consider the ground state to be excited by a rotating saddle potential
	\begin{equation}
		\hat{V}= \lambda_Q \left(\hat{O}_{x^2-y^2} \cos(\Omega t) + \hat{O}_{xy} \sin(\Omega t)\right)e^{-t^2/\tau^2},
	\end{equation}
	where
	\begin{align}
		\hat{O}_{x^2-y^2} &= \sum_i (x_i^2-y_i^2)\,\hat{n}_i
		\\
		\hat{O}_{xy} &= \sum_i 2x_i y_i\,\hat{n}_i.
	\end{align}
	The perturbation can be rewritten in terms of the complex combination $z_i=x_i+i y_i$ as
	\begin{equation}
		\label{eq:rotating_saddle}
		\hat{V}= \frac{1}{2}\lambda_Q \left(\hat{O}_{z^2}^\nodagger e^{-i\Omega t} + \hat{O}_{z^2}^\dagger e^{+i\Omega t}\right)e^{-t^2/\tau^2}
	\end{equation}
	with $\hat{O}_{z^2} = \sum_i z_i^2\,\hat{n}_i = \hat{O}_{x^2-y^2} + i \hat{O}_{xy}$.
	
	If the system was fully rotationally symmetric, due to angular momentum conservation we would have $\braket{q_1|\hat{O}_{z^2}^\dagger|0}=\braket{q_2|\hat{O}_{z^2}^\dagger|0}=0$. 
	However, our system has only fourfold rotational symmetry. As a consequence, these matrix elements do not vanish, even though we find them to be significantly smaller (by a factor $\sim10$). We show a comparison in Fig.~\ref{fig:quadrupoles}.
	The purpose of the modulation at frequency $\Omega$ is to make transitions driven by $\hat{O}_{z^2}^\dagger$ from $\ket{0}$ to the quadrupole modes $\{\ket{q_1}, \ket{q_2}\}$ small; this allows one to focus instead on the transitions driven by $\hat{O}_{z^2}^\nodagger$, of which we want to reconstruct the static structure factor.
	
	Using linear-order perturbation theory we write
	\begin{equation}
		\label{eq:time_evo_quadrupole}
		Q_z(t)\simeq \frac{i}{\hbar}\int_{-\infty}^t 
		\Braket{\left[\widetilde{V}(t'), \widetilde{O}_{z^2}(t)\right]}\,dt'
	\end{equation}
	where the tilde denotes Heisenberg representation with respect to $\hat{H}$ and we defined $Q_z(t)=\braket{\hat{O}_{z^2}}(t)$. 
	Notice that $\hat{O}_{z^2}(t)=\hat{O}_{x^2-y^2}+i \hat{O}_{xy}$ is a complex linear combination of two physical observables {\color{black} which transforms as a $l=2$ angular momentum operator under rotations}.
	
	Neglecting off-resonant terms in Eq.~\eqref{eq:time_evo_quadrupole} we obtain
	\begin{align}
		Q_z(t)=\frac{i}{2\hbar}\int_{-\infty}^t dt'\lambda(t')\sum_n \left( e^{i \omega_{n,0}t} e^{i(\Omega-\omega_{n,0})t'}|\braket{n|\hat{O}_{z^2}^\nodagger|0}|^2
		- 
		e^{-i \omega_{n,0}t} e^{-i(\Omega-\omega_{n,0})t'} \braket{0|\hat{O}_{z^2}^\nodagger|n}\braket{n|\hat{O}_{z^2}^\nodagger|0}
		\right).
	\end{align}
	For large enough $\tau$ and $\Omega\simeq \omega_q$ only one of the quadrupole modes $\ket{q}$ will be excited. For $t\gg\tau$ we write
	\begin{align}
		\label{eq:perturbativeQuadrupole}
		Q_z(t\gg\tau)=\frac{i}{2\hbar} \left( e^{i \omega_q t}  \widetilde{\lambda}(\omega_q-\Omega) |\braket{q|\hat{O}_{z^2}^\nodagger|0}|^2
		- 
		e^{-i \omega_q t} \widetilde{\lambda}^*(\omega_q-\Omega) \braket{0|\hat{O}_{z^2}^\nodagger|q}\braket{q|\hat{O}_{z^2}^\nodagger|0}
		\right).
	\end{align}
	{\color{black} 
		Positive and negative frequency components have different amplitudes: indeed,
		given the condition between the matrix elements of $\hat O_{z^2}$ and $\hat O_{{z^*}^2}$ we discussed above, the term oscillating as $e^{-i\omega_q t}$ has a considerably smaller amplitude. 
		As a consequence, the real and imaginary parts of $Q_z(t)$ (i.e. the expectation value of $\hat O_{x^2-y^2}$ and $\hat O_{xy}$) will oscillate (out-of-phase) with different amplitudes, albeit at the same frequency. This can be appreciated in Fig.~3(c) of the main text.
		Physically this difference in oscillation amplitude can be understood in terms of the geometry of the square lattice confinement, which is effectively stiffer (softer) in the Cartesian (diagonal) directions associated to the $x^2-y^2$ ($xy$) components of the quadrupole.}
	By resolving the two frequencies, for example through Fourier analysis, the relevant amplitude $|\braket{q|\hat{O}_{z^2}^\nodagger|0}|^2$ can be extracted; 
	once this is done for the first quadrupole mode $\ket{q_1}$, 
	the same can be done for the second one, $\ket{q_2}$
	and one can reconstruct the static structure factor and match it with Wen's prediction.
	We show the discrete Fourier transforms (taken only for $t\gtrsim\tau$) of $Q_z(t)$ for the two quadrupolar transitions in Fig.~\ref{fig:quadrupole_powerSpectrum}. Frequencies are measured in units of the relevant transition frequency $\omega_q$.
	Peaks at $\omega/\omega_q=\pm1$ can be seen; as expected, the one at negative frequency, which emerges because of the reduced rotational symmetry of the lattice, can be seen to be smaller than the relevant one at positive frequency.
	The peak-amplitude of the latter is compared with the amplitude that we obtain from the perturbation theory result (yellow-dashed line) and with the non-linear two-level system result Eq.~\eqref{eq:rabi_quadrupole}.
	It can be seen that the latter compares successfully with the numerical results and can be used to estimate the relevant matrix element $|\braket{q|\hat{O}_{z^2}|0}|^2$.
	
	\begin{figure}[t]
		\centering
		\includegraphics[width=10cm]{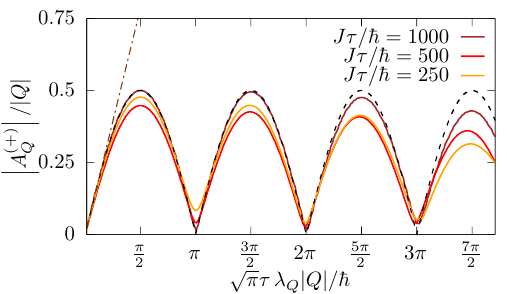}
		\caption{Positive frequency amplitude of $Q_z(t\gg\tau)$ as a function of the pulse angle $\Omega_R T$ ($T=\sqrt{\pi}\tau$), for different values of the pulse duration $\tau$. 
			The resonant two-level Rabi oscillation prediction (see Eq.~\eqref{eq:rabi_quadrupole}) is shown as a black-dashed line, and the linear approximation (see Eq.~\eqref{eq:perturbativeQuadrupole}) is shown as a dash-dotted brown line. 
			{\color{black}The system consists here of $N=3$ hard-core bosons hopping on a $9\times9$ lattice, at synthetic magnetic flux per plaquette $\alpha=0.15$. The targetted quadrupole state is the one with $\omega_q=0.035\hbar/J$.}
			\label{fig:rabiOscillations_Q1}}
	\end{figure}
	\subsection{Rabi oscillations}
	Analogously to the dipolar-excitation case, due to the strong non-linearities imposed by the lattice the ground-state to quadrupole transition can form an isolated two-level manifold within the exponentially large set of many-body states.
	The quantitative analysis is the same as in the case of the dipolar excitation, with the additional complication of the reduced rotational symmetry of the lattice.
	We find, for a resonant excitation ($\Omega=\omega_q$)
	\begin{equation}
		\label{eq:rabi_quadrupole}
		Q_z(t\gg\tau)=i\,\underbrace{\frac{\left|Q\right|}{2}\sin\left(\sqrt{\pi}\tau\,\frac{\lambda_Q\left|Q\right|}{\hbar}\right)}_{A_Q^{(+)}}\,e^{i\omega_q t} -i \underbrace{\frac{Q'e^{i\alpha}}{2}\sin\left(\sqrt{\pi}\tau\,\frac{\lambda_Q\left|Q\right|}{\hbar}\right)}_{A_Q^{(-)}}\,e^{-i\omega_q t}
	\end{equation}
	where $Q=\braket{q|\hat{O}_{z^2}^\nodagger|0}$, $Q'=\braket{0|\hat{O}_{z^2}^\nodagger|q}=(\braket{q|\hat{O}_{{z}^2}^\dagger|0})^*$ and $\alpha=\arg\left(Q\right)$ {\color{black}(see Fig.~\ref{fig:quadrupoles} for a plot of these matrix elements)}.
	To linear order in perturbation theory, this result reduces to the perturbative one given in Eq.~\eqref{eq:perturbativeQuadrupole}.
	
	We compare the numerical amplitude of the peak at positive frequency, $A_Q^{(+)}$, with the above result Eq.~\eqref{eq:rabi_quadrupole} in Fig.~\ref{fig:rabiOscillations_Q1}, for the case of the lower quadrupolar transition. It can be seen that for long-enough pulses the predictions nicely match the analytical result.
	
	\begin{figure}[t]
		\centering
		\includegraphics[width=17.2cm]{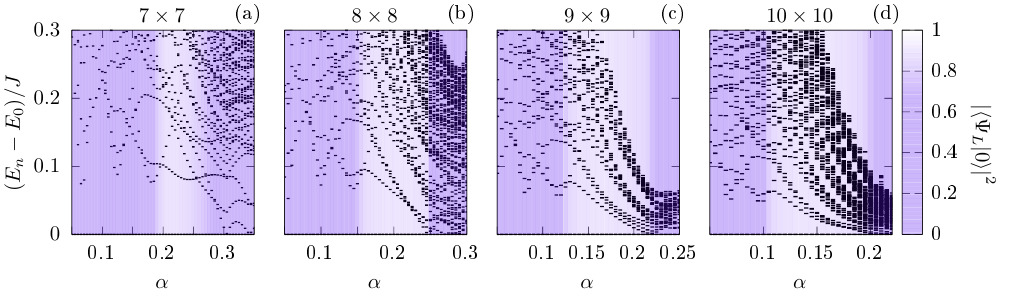}
		\caption{Energy levels $E_n$ (with respect to the system's ground state energy $E_0$) as a function of the magnetic flux per plaquette $\alpha$, for $N=4$ hard-core bosons moving in a (a) $7\times7$, (b) $8\times8$, (c) $9\times9$, (d) $10\times10$ square lattice.
			The purple background transparency is regulated according to the (squared) overlap of the ground state $\ket{0}$ with the discretized Laughlin state of Eq.~\eqref{eq:Laughlin}, $\ket{\Psi_L}$.
			In order for the data point not to disappear when the overlap drops, when the squared overlap drops below $0.7$ the transparency is kept constant.
			\label{fig:spectra_4vsM}}
	\end{figure}
	
	\begin{figure}[t]
		\centering
		\includegraphics[width=17.2cm]{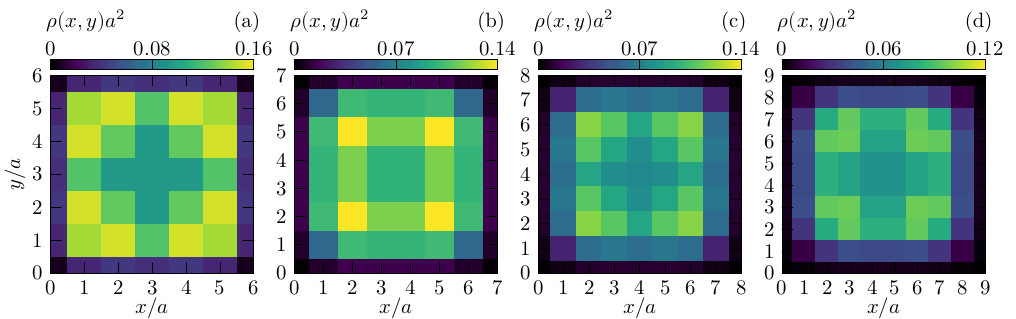}
		\caption{Ground state densities of the $N=4$ hard-core boson system in lattices with different sizes, for some selected values of the magnetic flux per plaquette $\alpha$ which correspond to a ground state in the Laughlin phase.
			In particular, the four panels correspond to
			(a) $7\times7$ lattice with $\alpha = 0.22$,
			(b) $8\times8$ lattice with $\alpha = 0.21$,
			(c) $9\times9$ lattice with $\alpha = 0.17$ and
			(d) $10\times10$ lattice with $\alpha = 0.15$. 
			The particular values of $\alpha$ have been chosen so as to (approximately) maximize the overlap (see Fig.~\ref{fig:overlaps}) of the given state with the discretized Laughlin wavefunction in Eq.~\eqref{eq:Laughlin}.
			\label{fig:densities}}
	\end{figure}
	
	\section{Additional numerical simulations}
	
	\subsection{Lattice-size effects}
	In this section we analyze how the number of lattice sites influences the energy spectra and the structure factors.
	
	\noindent In Fig.~\ref{fig:spectra_4vsM} we show the excitation spectra for $N=4$ particles on a $M\times M$ lattice, for $M=7,\hdots,10$.
	The background transparency is regulated according to the overlap of the ground state with a discretized bosonic Laughlin wavefunction
	\begin{equation}
		\label{eq:Laughlin}
		\Psi_L=\prod_{1\leq i<j\leq N} (z_i-z_j)^{1/\nu}\,e^{-\frac{1}{4l_B^2}\sum_{i=1}^N |z_i|^2}.
	\end{equation}
	Here $\nu=1/2$ is the (robust) filling fraction associated to the particular  incompressible phase, while $z_i=x_i+i y_i$ take values on the square lattice.
	A large region in which the ground state has a large overlap with this model-wavefunction can be seen (see also Fig.~\ref{fig:overlaps}).
	
	{\color{black} From classical considerations the velocity of the edge modes is expected to scale as $\approx E/B$, $E$ being the effective (synthetic) electric field providing confinement for the particles and $B$ the (synthetic) magnetic field.}
	{\color{black} The larger the box, the farther away the atoms will be from} the boundary lattice sites (see Fig~\ref{fig:densities}) {\color{black} and thus the weaker the effective confining field $E$.}
	It can {\color{black} indeed} be seen from Fig.~\ref{fig:spectra_4vsM} that as the size of the lattice is increased (from the leftmost panel to the rightmost), the edge-excitation velocity decreases {\color{black}(the energy of the dipole mode with respect to the Laughlin one decreases)}. {\color{black} According to these considerations, we also highlight how  the velocity of the edge modes decreases as the synthetic flux per plaquette $\alpha$ is increased.}
	
	\begin{figure}[t]
		\centering
		\includegraphics[width=17.2cm]{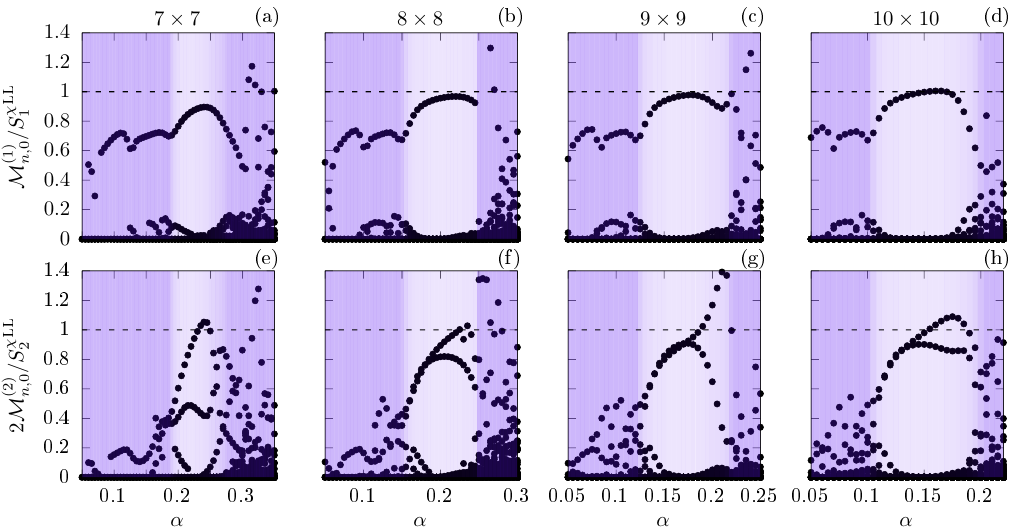}
		\caption{
			Matrix elements $\mathcal{M}_{n,0}^{(l)}=\left|\Braket{n|\sum_i z_i^l \hat{n}_i|0}\right|^2$ as a function of the magnetic flux per plaquette $\alpha$, for $N=4$ hard-core bosons, moving in a (a,e) $7\times7$, (b,f) $8\times8$, (c,g) $9\times9$, (d,h) $10\times10$ square lattice.
			The top (a-d) (/bottom (e-h)) panels are for $l=1$ (/$l=2$).    
			The purple background transparency is regulated according to the (squared) overlap of the ground state $\ket{0}$ with the discretized Laughlin state of Eq.~\eqref{eq:Laughlin}, $\ket{\Psi_L}$.
			In order for the data point not to disappear when the overlap drops, when the squared overlap drops below $0.7$ the transparency is kept constant.
			The static structure factor per state $S_l^\cll/l$ given by Eq.~\eqref{eq:ssf} is shown as black dashed lines.    
			\label{fig:structure_factor_4vsM}}
	\end{figure}
	
	In Fig.~\ref{fig:structure_factor_4vsM} we then plot the matrix elements $\mathcal{M}_{n,0}^{(l)}$ for $l=1,2$; 
	since in the Laughlin region only $l$ modes have large matrix-elements, we normalize the results to the integrated structure factor per state in the continuum, $S^\cll_l / l = \nu R_\text{cl}^{2l}$, $R_\text{cl}=\sqrt{2N/\nu} l_B$ being the classical radius and $l_B$ the magnetic length. 
	This result can be obtained from a \cll description of the edge excitations of a Laughlin ground state in the continuum, as we explicitly showed in the second section.
	With such a normalization, summing over the $l$ non-vanishing matrix elements should therefore give $l$, provided the geometric factor $R_\text{cl}^{2l}$ is not strongly affected by the shape of the square boundary.
	As the system size is increased, it can be seen in Fig.~\ref{fig:structure_factor_4vsM}(a-d) that the dipole matrix element ($l=1$) does indeed approach the expected result $S^\cll_1 = \nu R_\text{cl}^{2}$; 
	panels(e-h) show instead what happens for the quadrupole matrix elements: 
	in the Laughlin region (highlighted by the purple transparency)
	as the number of lattice sites is increased two states only (full circles) contribute to the structure factor $S^\cll_2$; 
	it can also be seen how their magnitude approaches $S^\cll_2 / 2 = \nu R_\text{cl}^{4}$, hinting to the correctness of the sum rule Eq.~\eqref{eq:approximated_SF}. 
	We will show in greater detail in the next section (in particular in Fig.~\ref{fig:static_structure_factor_N}) how $S^\cll_2$ changes as a function of the magnetic flux per plaquette $\alpha$, for various lattice sizes and number of particles.
	
	\begin{figure}[ht!]
		\centering
		\includegraphics[width=17.2cm]{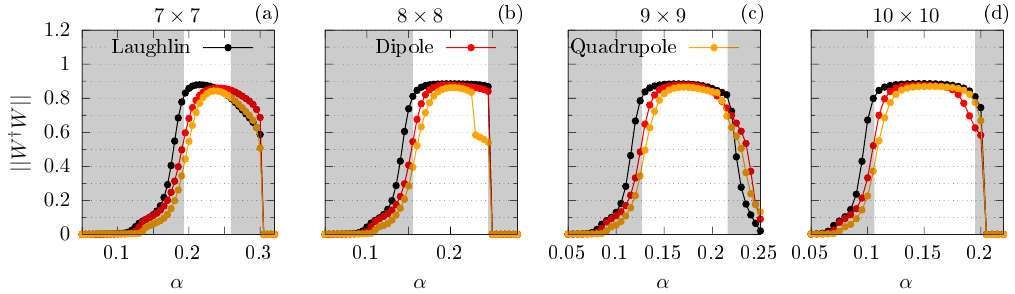}
		\caption{Overlaps between selected numerical eigenstates and model fractional quantum Hall wavefunctions for the Laughlin state Eq.~\ref{eq:Laughlin} and its edge excitations Eq.~\eqref{eq:dipole}, Eq.~\eqref{eq:quadrupolar},
			for a $N=4$ system of hard-core bosons moving on
			(a) $7\times7$,
			(b) $8\times8$,
			(c) $9\times9$ and
			(d) $10\times10$ lattices, as a function of the synthetic magnetic flux per plaquette. 
			{\color{black}As a guide to the eye, a white transparency window has been drawn to signal that the absolute value of the overlap between the ground state and the discretized Laughlin state is larger than $0.9$ (notice that in the case of the Laughlin state, the black curves correspond to the {\it squared} overlap and not to its absolute value).}
			\label{fig:overlaps}}
	\end{figure}
	
	Finally, in Fig.~\ref{fig:overlaps} we plot the overlaps of a few selected eigenstates with discretized model fractional quantum Hall wavefunctions for the gapless edge excitations of a Laughlin liquid, for different system sizes.
	In particular, the ground state, with symmetry eigenvalue $C_4$, which we label $\ket{C_4,0}$, is compared to the discretized Laughlin wavefunction of Eq.~\eqref{eq:Laughlin}; 
	a good agreement (black curves) for the squared overlaps ($|\Braket{\Psi_L|C_4,0}|^2$ up to $\sim0.9$) can be seen over a wide range of synthetic magnetic flux per plaquette $\alpha$, for various lattice sizes.
	The lowest lying excitation $\ket{C_4',0}$, where $C_4'=C_4+1 \pmod 4$,  is compared instead to the edge excitation
	\begin{equation}
		\label{eq:dipole}
		\Psi_d = \sum_i z_i \prod_{1\leq i<j\leq N} (z_i-z_j)^{1/\nu}\,e^{-\frac{1}{4l_B^2}\sum_{i=1}^N |z_i|^2}.
	\end{equation}
	Again, a good agreement (red curves) can be found in the same region of $\alpha$, although this window is slightly narrower.
	
	Finally, the two quadrupolar modes $\ket{C_4'',0}$ and $\ket{C_4'',1}$ can be compared to discretized quadrupolar excitations of a Laughlin state
	\begin{equation}
		\label{eq:quadrupolar}
		\begin{split}
			\Psi_q &= \left(\sum_i z_i\right)^2 \prod_{1\leq i<j\leq N} (z_i-z_j)^{1/\nu}\,e^{-\frac{1}{4l_B^2}\sum_{i=1}^N |z_i|^2}\\
			\Psi_q' &=\sum_i z_i^2 \prod_{1\leq i<j\leq N} (z_i-z_j)^{1/\nu}\,e^{-\frac{1}{4l_B^2}\sum_{i=1}^N |z_i|^2}.
		\end{split}
	\end{equation}
	These two states $\ket{\Psi_q}$ and $\ket{\Psi_q'}$ are linearly independent but not orthonormal:
	we therefore orthonormalize them with a standard Gram-Schmidt procedure; moreover, it does not make sense to compare the two separately to the two quadrupolar states $\ket{C_4'',0}$ and $\ket{C_4'',1}$:
	in principle we can however expect these to be a mixture of $\ket{\Psi_q}$ and $\ket{\Psi_q'}$ which minimizes the particular Hamiltonian, provided mixing with states outside of this manifold is small. 
	We test this idea by defining a matrix of overlaps
	\begin{equation}
		W =
		\begin{pmatrix}
			\braket{\Psi_q|C_4'',0} & \braket{\Psi_q'|C_4'',0}\\
			\braket{\Psi_q|C_4'',1} & \braket{\Psi_q'|C_4'',1}
		\end{pmatrix}
	\end{equation}
	and testing whether it defines an unitary matrix (i.e. a ``rotation" between two otherwise equivalent bases) by computing an Hilbert-Schmidt norm of $W^\dagger W^\nodagger$
	\begin{equation}
		\label{eq:HSNorm}
		||W^\dagger W^\nodagger|| = \sqrt{\frac{1}{d}\sum_{i,j=1}^d \left|(W^\dagger W^\nodagger)_{i,j}\right|^2},
	\end{equation}
	where $d$ is the dimension of the subspace under consideration; in this case, $d=2$.
	Ideally, $W^\dagger W^\nodagger=\mathbb{1}$ and the normalization in Eq.~\eqref{eq:HSNorm} has been chosen so that $||\mathbb{1}||=1$.
	Notice finally that when $d=1$, Eq.~\eqref{eq:HSNorm} reduces to the squared overlap used in the two previous cases with the ground- and dipole- states.
	Again, a good agreement (yellow curve) can be seen in the relevant region of $\alpha$ in Fig.~\ref{fig:overlaps}, with $||W^\dagger W^\nodagger||\approx0.9$.
	
	Although these overlaps can be expected to drop to zero in the thermodynamic limit~\footnote{As it is usually the case for the overlaps between model wavefunctions and “true” ground-state wavefunctions, the overlap can be expected to go to zero as the number of particles is increased.} ($N,L\rightarrow\infty$), the fact that they are significantly close to $1$ in a limited window of fluxes per plaquette for the lattice sizes we analysed supports our interpretation of them being the edge modes of a Laughlin state.
	Let us finally stress that the fact that we expect these overlaps to drop does not mean that our conclusions will be altered: on the contrary, Eq.~\eqref{eq:approximated_SF} is a thermodynamic limit statement and a consequence of dealing with a Laughlin topological order~\footnote{Namely, it does not rely on the assumption of dealing with a Laughlin wavefunction, but rather on having a single chiral edge-mode satisfying Eq.~\eqref{eq:wen_commutator}.}, and not of exactly having model Laughlin-like wavefunctions for the ground state and its excitation.
	
	\begin{figure}[ht!]
		\centering
		\includegraphics[width=17.2cm]{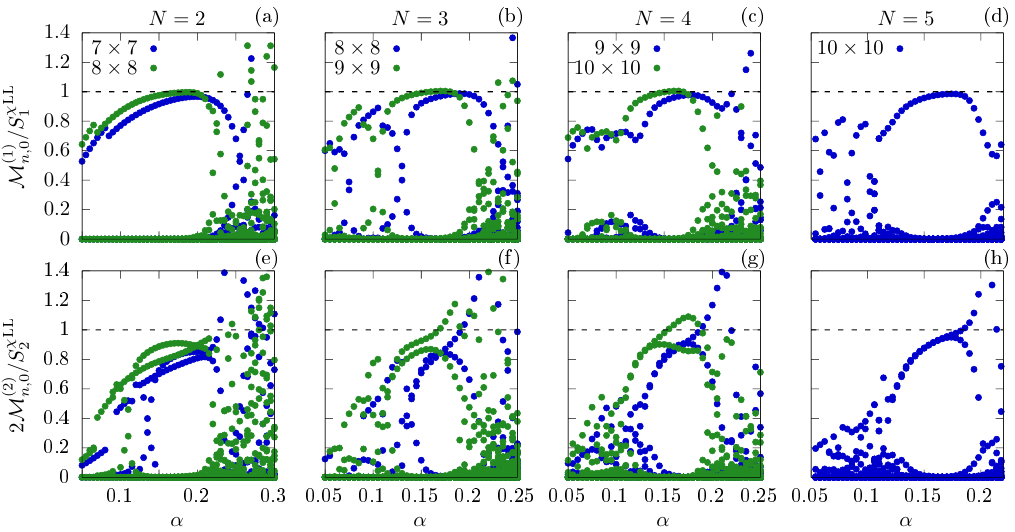}
		\caption{
			Matrix elements $\mathcal{M}_{n,0}^{(l)}=\left|\Braket{n|\sum_i z_i^l \hat{n}_i|0}\right|^2$ as a function of the magnetic flux per plaquette $\alpha$, for (a,e) $N=2$ (b,f) $N=3$, (c,g) $N=4$ and (d,h) $N=5$ hard-core bosons, for different lattice sizes.
			The top (a-d) (/bottom (e-h)) panels are for $l=1$ (/$l=2$).
			The static structure factor per state $S_l^\cll/l$ given by Eq.~\eqref{eq:ssf} is shown as black dashed lines.
			\label{fig:structure_factor_N}}
	\end{figure}
	
	\begin{figure}[ht!]
		\centering
		\includegraphics[width=17.2cm]{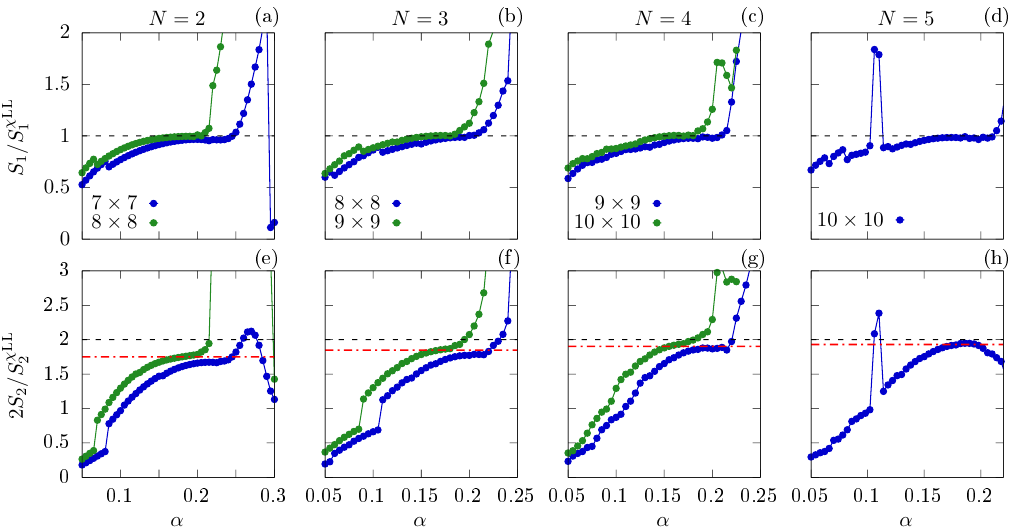}
		\caption{
			Static structure factors $S_l = \sum_n \mathcal{M}_{n,0}^{(l)}$ as a function of the magnetic flux per plaquette $\alpha$, normalized to the static structure factor per contributing state $S_l^\cll/l$ given by Eq.~\eqref{eq:wen_SF};
			in the various panels, different system sizes are shown: (a,e) $N=2$ (b,f) $N=3$, (c,g) $N=4$ and (d,h) $N=5$.
			The expected static structure factor given by Eq.~\eqref{eq:wen_SF} is shown as black dashed lines.
			The top (a-d) (/bottom (e-h)) panels are for $l=1$ (/$l=2$).
			In the $l=2$ cases (panels (e-h)), the red dashed-dotted lines are the results for the static structure factor Eq.~\eqref{eq:ssf_rewriting} for a continuum state (see also Fig.~\ref{fig:continuum}).
			\label{fig:static_structure_factor_N}}
	\end{figure}
	
	\subsection{Scaling with the number of particles}
	In this section we briefly show and discuss the results for the matrix elements $\mathcal{M}_{n,0}^{(l)}$ for different number of particles.
	
	In particular, in Fig.~\ref{fig:structure_factor_N} the results in the $l=1,2$ cases are shown as a function of the synthetic magnetic flux per plaquette $\alpha$, for different lattice sizes. 
	In the top panels (a-d) we show the results for $\mathcal{M}_{n,0}^{(1)}$. It can be seen that 
	a single state has a significantly non-zero matrix element in the region of $\alpha$ in which the ground state is a Laughlin state. 
	Only this state, which we identify with the dipole excitation of Eq.~\eqref{eq:dipole}, contributes to the integrated structure factor of Eq.~\eqref{eq:ssf}; indeed it can be seen that in the Laughlin region the value of this matrix element collapses to the static structure factor curve Eq.~\eqref{eq:wen_SF} (black dashed line), for every $N$ and $L$ displayed here.
	
	In the bottom panels (e-h) we instead show the results for $\mathcal{M}_{n,0}^{(2)}$. It can be seen that 
	only two states contribute significantly to the static structure factor Eq.~\eqref{eq:ssf}; we interpret these two states as the quadrupole modes of a Laughlin state, Eq.~\eqref{eq:quadrupolar}.
	The matrix elements have been compared to the expected static structure factor per state, $S_l^\cll/l$ (black dashed line); 
	as we already discussed in the previous section, it is important to notice that there is no a priori reason why the two matrix elements should be equally contributing to $S_l^\cll$: in principle only one of them could be carrying all the spectral weight (see for example Eq.~\eqref{eq:approximated_SF}, Eq.~\eqref{eq:rho_l} and Eq.~\eqref{eq:state} above).
	It can however be seen, at least qualitatively, that as the number of particles is increased, the expected value of $S_l^\cll$ is reached.
	
	We also performed a more quantitative test. In particular, in Fig.~\ref{fig:static_structure_factor_N} we show the integrated dynamic structure factor, namely $S_l=\sum_n \mathcal{M}_{n,0}^{(l)}$. In the $l=1$ case (panels (a-d)) a good agreement with the prediction for $S_l^\cll$ given by Wen's theory (black dashed line), Eq.~\eqref{eq:wen_SF}, can be seen.
	In the $l=2$ case (panels e-h)), even though the integrated structure factor $S_l$ deviates from this prediction, as the number of particles is increased the plateau that $S_l$ exhibits gets closer to $S_l^\cll$. 
	Furthermore, this finite size correction is compatible with the results (red dotted dashed line) that we obtained for a Laughlin state in the continuum (see Fig.~\ref{fig:continuum}).
	
	\begin{figure}[ht!]
		\centering
		\includegraphics[width=14cm]{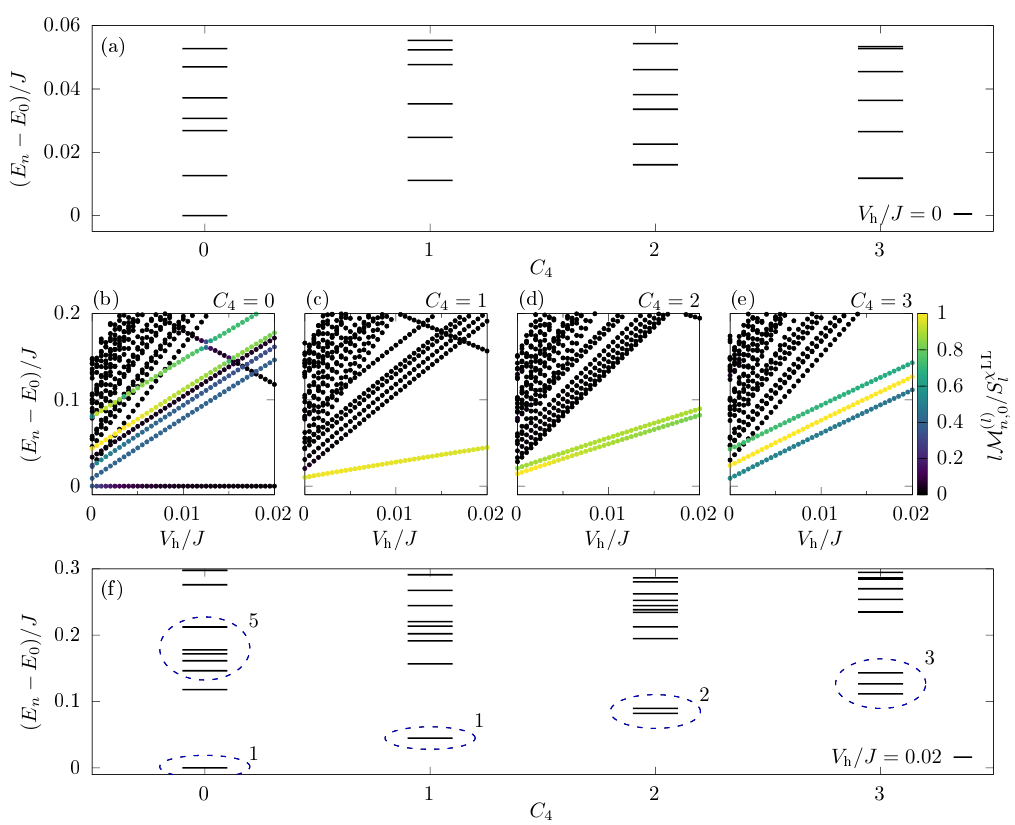}
		\caption{(a) Energy levels (with respect to the ground state one) of a system of $N=4$ hard-core bosons moving on a $10\times10$ lattice, at $\alpha=0.18$. The energies are plotted against the fourfold discrete rotation symmetry eigenvalue $C_4$.
			\newline (b-e) Energy levels (with respect to the ground state one) for the same system, as a function of the harmonic confinement strength $V_\text{h}$ (see Eq.~\eqref{eq:harmonic_confinement}; the various panels refer to the different $C_4$ values: (b) $C_4=0$ (c) $C_4=1$ (d) $C_4=2$ and (e) $C_4=3$.
			The points are coloured according to the value of the matrix elements $\mathcal{M}_{n,0}^{(l)}=\left|\braket{n\left|\sum_i z_i^l \hat{n}_i\right|0}\right|^2$, for (b) $l=4$, (c) $l=1$, (d) $l=2$ and (e) $l=3$.
			\newline (f) Energy levels (with respect to the ground state one) plotted against the symmetry eigenvalue $C_4$ for the same system, in the presence of an additional harmonic confinement with strength $V_h/J=0.02$. Dashed-blue ellipses highlight the now well isolated group of edge modes with the counting predicted for a Laughlin ground state~\cite{Kjall_PRB_2012,Binanti_PRR_2024}.
			\label{fig:harmonic}}
	\end{figure}
	
	\subsection{Effect of additional harmonic confinement}
	In this section we analyse the effect of an additional harmonic confinement which does not spoil the fourfold rotational symmetry of the lattice; namely, we add to the Harper-Hofstadter Hamiltonian an additional potential term
	\begin{equation}
		\label{eq:harmonic_confinement}
		\hat{V}_\text{conf} = \sum_{i,j} V_\text{h} a^2\left( \left(i-\frac{M-1}{2}\right)^2 + \left(j-\frac{M-1}{2}\right)^2 \right) \hat{n}_{ij}.
	\end{equation}
	We study the energies and matrix elements $\mathcal{M}_{n,0}^{(l)}$ as a function of $V_\text{h}$ in Fig.~\ref{fig:harmonic}.
	
	In the absence of the harmonic confinement, when plotted against the fourfold discrete rotation symmetry eigenvalue $C_4$, 
	the level structure cannot be easily interpreted visually (see Fig.~\ref{fig:harmonic}(a)) due to the competing energy-scales of bulk and edge excitations.
	However, a weak harmonic confinement {\color{black}reinforces the group velocity of the edge excitations and, in this way,} isolates in each symmetry sector a low-lying group of states with the counting predicted by the edge-mode theory for a Laughlin state (see Fig.~\ref{fig:harmonic}(f)).
	The degeneracy is slightly shifted by perturbations of Wen's Hamiltonian, leading to a weakly non-linear \cll~\cite{Fern_PRB_2018,Nardin_PRA_2023,nardin2023refermionized}.
	{\color{black} Compared to the sharp boundaries of open lattices, the effective confinement in the presence of an additional harmonic one is smoother and therefore reduces the relative importance of non-linear effects.}
	
	\begin{figure}[t]
		\centering
		\includegraphics[width=14cm]{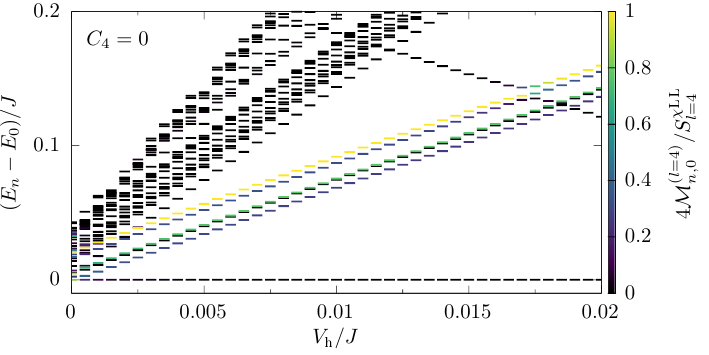}
		\caption{Same as Fig.~\ref{fig:harmonic}(b), but for an $11\times11$ lattice.
			\label{fig:harmonic2}}
	\end{figure}
	
	In the mid panels, Fig.~\ref{fig:harmonic}(b-e), we show that the ground state in the absence of the harmonic confinement Eq.~\eqref{eq:harmonic_confinement} is adiabatically connected to the ground state in its presence, which is unambigously identified to be a Laughlin topological order by the edge-mode counting seen in Fig.~\ref{fig:harmonic}(f). 
	Furthermore, the lowest lying states at $l=1$ and $l=2$ can be seen to be connected with the low-lying edge excitations highlighted in Fig.~\ref{fig:harmonic}(f).
	This supports the results we previously exhibited in Fig.~\ref{fig:overlaps} for the overlaps with the discretized model wavefunctions for the Laughlin state of Eq.~\eqref{eq:Laughlin} and its edge excitations Eq.~\eqref{eq:dipole} and Eq.~\eqref{eq:quadrupolar}.
	
	The points in Fig.~\ref{fig:harmonic}(b-e) are also coloured according to the matrix elements $\mathcal{M}_{n,0}^{(l)}=\left|\braket{n\left|\sum_i z_i^l \hat{n}_i\right|0}\right|^2$; 
	it is worth highlighting some features.
	First of all in the $C_4=3$ plot in Fig.~\ref{fig:harmonic} (panel (e)) it can be seen that the $3$ states that carry non-zero weight for the $l=3$ excitation
	are those that at large $V_\text{h}$ can be identified with the $3$ edge excitations of a continuum Laughlin liquid at the same angular momentum variation with respect to the ground state; at small $V_\text{h}$ these states cross with other levels, but they do not mix with them.
	Secondly, for $N=4$ particles, also the counting of edge modes in the $C_4=4\equiv0\pmod4$ symmetry sector shown Fig.~\ref{fig:harmonic}(b) is expected to show the thermodynamic-limit counting~\cite{Wen_AdvPhys_1995,SimonWavefunctionology_2020} (i.e. a group of $5$ states). 
	For the parameters we used however this is not so clear.
	Indeed in Fig.~\ref{fig:harmonic}(b) it can be seen that around $V_\text{h}/J=0.01$ there is a low-lying group of $5$ states which couple to the density operator in the same way a non-harmonically confined Laughlin state does~\cite{Nardin_PRA_2023,nardin2023refermionized} - namely only $4$ states couple to it while the fifth one is dark;
	however, as $V_\text{h}$ is further increased there is a level crossing (for which we lack an intuitive understanding) at around $V_\text{h}/J\approx0.12$, mixing in this subspace.
	This makes the interpretation of the low-lying group of $5$ states at $V_\text{h}/J=0.02$ (seen both in Fig.~\ref{fig:harmonic}(b) and (f)) a bit ambiguous.
	A well isolated group of $5$ states can indeed be obtained (in the region $0.005\lesssim V_\text{h}/J\lesssim0.015$) by slightly increasing the size of the lattice, as we show in Fig.~\ref{fig:harmonic2}. Also in this case it can be seen that one of the five edge states is dark (which may be difficult to spot at a fast glance because ``hidden" below the green coloured points), analogously to the weakly non-linear Luttinger liquid~\cite{Nardin_PRA_2023,nardin2023refermionized}.
	
	It is crucial here to notice that, even though removing the harmonic confinement (as $V_\text{h}$ is decreased) introduces many level crossings of the edge-modes, the eigenvectors do not mix;
	even though the energetic structure of the \cll is lost, the eigenvectors survive these strong perturbations which do not close the bulk many-body gap. 
	Roughly speaking, since the topological order does not change, the structure of the edge theory must still be the same - and carry the information of the quantized transverse Hall current.
	
	As a final minor comment, notice in Fig.~\ref{fig:harmonic}(b) that $\mathcal{M}_{0,0}^{(4)}$ is non-zero when $V_\text{h}\simeq0$, as can also be seen in Fig.~2 of the main text. For the parameters of Fig.~\ref{fig:harmonic}(b), at $V_h=0$, $4\mathcal{M}_{0,0}^{(4)} / S_4^\cll\simeq 0.38$.
	This non-zero {\color{black} static octupole moment of the ground state density distribution} is allowed by the reduced four-fold symmetry of the lattice.
	We notice here however that $\mathcal{M}_{0,0}^{(4)}$ goes to zero as the strength of the harmonic confinement is increased, approaching thus the value expected in the continuum from angular momentum conservation; namely, {\color{black} this quantity tends to zero when the confinement starts being dominated by the harmonic potential (which is cylindrically symmetric) rather than by the square one naturally provided by the lattice open boundaries.}
	
\end{supplementalMaterials}
	
\end{document}